\documentclass[a4paper,10 pt, journal]{irepclass}

\usepackage[latin1]{inputenc}
\usepackage[T1]{fontenc}
\usepackage{lmodern}
\usepackage{amsmath}
\usepackage{amssymb}
\usepackage{color}
\usepackage{graphics}
\usepackage{epsfig}
\usepackage{epsf}
\usepackage{graphicx}
\usepackage{multirow}
\usepackage{capt-of}
\usepackage{url}
\usepackage{fancybox}
\usepackage{eurosym}

\usepackage[parfill]{parskip}

\newenvironment{mydescription}[1]
  {\begin{list}{}%
   {\renewcommand\makelabel[1]{##1 \hfill}%
   \settowidth\labelwidth{\makelabel{#1}}%
   \setlength\leftmargin{\labelwidth}
   \addtolength\leftmargin{\labelsep}}}
  {\end{list}}

\usepackage{array}

\let\savehline\hline
\def\sphline{\noalign{\vskip3pt}\savehline\noalign{\vskip3pt}}

\begin{document}

\title{\textbf{ Whither probabilistic security management for real-time operation of power systems
?}}

\author{\large{Efthymios Karangelos\IEEEauthorrefmark{1},  Patrick Panciatici\IEEEauthorrefmark{2},  Louis Wehenkel\IEEEauthorrefmark{1}}\\
\IEEEauthorblockA{\IEEEauthorrefmark{1}\normalsize{University of Li\`{e}ge, Belgium}}
\IEEEauthorblockA{\IEEEauthorrefmark{2}\normalsize{RTE-DMA, France}}
}

\markboth{\small{2013 IREP S\MakeLowercase{ymposium}-B\MakeLowercase{ulk} P\MakeLowercase{ower} S\MakeLowercase{ystems} D\MakeLowercase{ynamics and} C\MakeLowercase{ontrol}-IX (IREP),  A\MakeLowercase{ugust} 25-30,2013, R\MakeLowercase{ethymnon}, C\MakeLowercase{rete}, G\MakeLowercase{reece} } } {\small{2013 IREP S\MakeLowercase{ymposium}-B\MakeLowercase{ulk} P\MakeLowercase{ower} S\MakeLowercase{ystems} D\MakeLowercase{ynamics and} C\MakeLowercase{ontrol}-IX (IREP),  A\MakeLowercase{ugust} 25-30,2013, R\MakeLowercase{ethymnon}, C\MakeLowercase{rete}, G\MakeLowercase{reece} } }

\maketitle

\section*{Abstract}
This paper investigates the stakes of introducing probabilistic approaches for the management of power system's security. In real-time operation, the aim is to arbitrate in a rational way between preventive and corrective control, while taking into account i) the prior probabilities of contingencies, ii) the possible failure modes of corrective control actions, iii) the socio-economic consequences of service interruptions. This work is  a first  step towards the construction of a globally coherent decision making framework for security management from long-term system expansion, via mid-term asset management, towards short-term operation planning and real-time operation.

\IEEEpeerreviewmaketitle

\section*{Nomenclature}

\subsubsection*{Indices}
\begin{mydescription}{$DP_j^k$~}
\item[$b$]{Index of corrective control behaviors.}
\item[$c$]{Index of contingencies.}
\item[$d$]{Index of  demands.}
\item[$g$]{Index of generating units.}
\item[$\ell$]{Index of transmission lines.}
\item[$n$]{Index of nodes.}

\end{mydescription}

\subsubsection*{Sets}
\begin{mydescription}{$DP_j^k$~}

\item[$\mathcal{D}_{n}$]{Set of  demands connected at node $n$.}
\item[$\mathcal{G}_{n}$]{Set of generating units connected at node $n$.}
\item[$\mathcal{N}_{b}$]{Set of corrective control behaviors.}
\item[$\mathcal{N}_c$]{Set of contingencies.}
\item[$\mathcal{N}_d$]{Set of  demands. }
\item[$\mathcal{N}_g$]{Set of generating units.}

\item[$\mathcal{N}_{\ell}$]{Set of transmission lines.}
\item[$\mathcal{N}_n$]{Set of nodes. }
\end{mydescription}

\subsubsection*{Parameters}
\begin{mydescription}{$DP_j^k$~}
\item[$c_g$]{Marginal generation cost of generating unit $g$.}
\item[$c^r_g$]{Marginal corrective re-dispatch cost of generating unit $g$.}
\item[$P_{g}^{max}$]{Capacity of generating unit $g$.}
\item[$P_{g}^{min}$]{Minimum stable generation of unit $g$.}
\item[$P_{g}^{-}$]{Ramp-down limit of generating unit $g$.}
\item[$P_{g}^{+}$]{Ramp-up limit of generating unit $g$.}
\item[$\Delta P_{g}^{e}$]{Emergency ramp-down limit of generating unit $g$.}
\item[$w_g$]{Disconnection severity coefficient of generating unit $g$.}
\item[$P_{d}^0$]{Load of demand $d$.}
\item[$v_d$]{Value of lost load of demand $d$.}
\item[$f_{\ell}^{max}$]{Capacity of transmission line $\ell$.}
\item[$X_{\ell}$]{Reactance of transmission line $\ell$.}
\item[$\beta_{n,\ell}$]{Element of the flow incidence matrix, taking a value of one if node $n$ is the sending node of line $\ell$, a value of minus one if node $n$ is the receiving node of line $\ell$, and a zero value otherwise.}
\item[$\pi_c$]{Probability of occurrence of contingency $c$.}
\item[$\pi_b$]{Probability of  realization  of corrective control behavior $b$.}
\item[$s_{max}$]{Severity threshold.}
\item[$a_{i}^c$]{Binary parameter taking a zero value if  component $\{i \in \mathcal{N}_g\cup \mathcal{N}_{\ell}\} $ is unavailable under contingency $c \ge 1$.}
\item[$\tau_c$]{Binary parameter taking a value of one if contingency $c \ge 1$ concerns the failure of a generating unit.}
\item[$M$]{A large constant.}
\end{mydescription}

\subsubsection*{Continuous Variables}

\begin{mydescription}{$DP_j^k$~}
\item[$P_{g}^{0}$]{Power output of generating unit $g$ under the pre-contingency state.}
\item[$P_{g}^{c}$]{Corrective control schedule of generating unit $g$ under contingency $c \ge 1$.}
\item[$\hat{P}_{g}^{c}(b)$]{Power output of generating unit $g$ at the terminal state following the occurrence of contingency $c \ge 1$, the realization of corrective control behavior $b$ and the application of emergency control actions.}
\item[$f_{\ell}^{0}$]{Power flowing through transmission line $\ell$ under the pre-contingency state.}
\item[$f_{\ell}^{c}(b)$]{Power flowing through transmission line $\ell$ under contingency $c \ge 1$ and corrective control behavior $b $.}
\item[$\hat{f}_{\ell}^{c}(b)$]{Power flowing through transmission line $\ell$ at the terminal state following the occurrence of contingency $c \ge 1$, the realization of corrective control behavior $b$ and the application of emergency control actions.}
\item[$\hat{P}_{d}^{c}(b)$]{Load of demand $d$ at the terminal state following the occurrence of contingency $c \ge 1$, the realization of corrective control behavior $b$ and the application of emergency control actions.}
\item[$\theta_{n}^0$]{Voltage angle at  node $n$ under the pre-contingency state.}
\item[$\theta_{n}^c(b)$]{Voltage angle at node $n$ under contingency $c \ge 1$ and corrective control behavior $b$.}
\item[$\hat{\theta}_{n}^c(b)$]{Voltage angle at node $n$ at the terminal state following the occurrence of contingency $c \ge 1$, the realization of corrective control behavior $b $ and the application of emergency control actions.}
\item[$\delta_{n}^c(b)$]{Slackness on the power balance at node $n$ under contingency $c \ge 1$ and corrective control behavior $b$.}
\item[$s^c(b)$]{Severity of the terminal state following the occurrence of contingency $c \ge 1$, the realization of corrective control behavior $b $ and the application of emergency control actions.}
\end{mydescription}

\textit{Note: All continuous variables are non-negative with the exception of the line flow variables, voltage angle variables, and slack variable $\delta_{n}^c(b)$ .}

\subsubsection*{Binary Variables}
\markboth{}{}
\begin{mydescription}{$DP_j^k$~}
\item[$\lambda_{\ell}^{c}(b)$]{Binary variable, taking a value of one if there is an overload in transmission line $\ell$ under contingency $c \ge 1$ and corrective control behavior $b \in N_{b}$.}
\item[$p_{\ell}^{c}(b)$]{Binary variable, taking a value of one only if the flow of transmission line $\ell$ is positive under contingency $c \ge 1$ and corrective control behavior $b \in N_{b}$.}
\item[$\gamma^{c}(b)$]{Binary variable, taking a value of one if the severity of the terminal state following the occurrence of contingency $c \ge 1$ and the realization of corrective control behavior $b \in \mathcal{N}_{b}$ is greater than the respective threshold.}
\item[$y_{g}^c(b)$]{Binary variable taking a value of one  if generating unit $g$ has to be disconnected under contingency $c \ge 1$ and corrective control behavior $b$.}

\end{mydescription}

\section{Introduction}

In today's power systems security management practice, the $N-1$ criterion is used (with slightly different interpretations in different control areas)  to express an acceptable level of security. This fundamental criterion is consistently considered  throughout overlapping decision horizons, ranging from long-term system expansion, via mid-term asset management, to short-term operation planning and real-time operation.

{The  ageing of the power system infrastructure and the  increasing penetration of renewable and dispersed generation presently induce new threats to the system security. At the same time, the potential incorporation of emerging smart grid technologies poses as a new opportunity. Such developments call more and more for the explicit consideration of uncertainties through the development of alternative security criteria and  more effective decision making frameworks \cite{wehenkel-pscc99,panciatici2010security,varaiya-2011,capitanescu2011pscc,capitanescu2011state,vrakopoulou2012probabilistic}. }

The idea of framing power systems security management in a probabilistic paradigm is appealing but not new. {Since the 1970's, several proposals to formulate probabilistic variants of the problem and to develop solutions via tractable algorithmic approximations have been presented} (see,  \textit{e.g.} \cite{patton1972probability1,patton1972probability2,patton1972probability3,bouff1,dai}). Yet, none of these proposals has gained  acceptance by TSOs and Regulators{, as}  the practical difficulties for using probabilistic methods instead of the $N-1$ criterion are multifold: i) data quality issues, ii) computational complexity limitations, iii) the allocation of security provision and service interruption costs among control areas and end-users, iv) {methodological limitations in the  assessment} of  the social benefit of moving away from the $N-1$ criterion.

Since security management is a multi-stage/multi-actor optimisation problem (covering the different time-horizons from long-term to short-term and ranging over multiple subsystems coupled horizontally or vertically) it is necessary to be able to appraise  (and then be able to adapt to) any suggested change in the decision making strategy adopted in any one of these sub-problems.

\subsection{Proposal}

In the present paper we focus on the latest decision horizon, namely real-time operation. In this context, we investigate the stakes of introducing a probabilistic approach  to arbitrate between preventive and corrective control alternatives.

Reliance on corrective control measures is inevitably growing in response to the aforementioned increasing stress and uncertainty in power system security management. Yet, as with every system component, the operation of corrective control is also characterized by a certain degree of uncertainty. The fact that corrective control may fail to operate as anticipated implies that the use of this resource does not fully nullify the possibility of realizing undesirable system states resulting in unacceptable service disruption levels. Explicitly acknowledging this feature, we propose to account not only for the operational cost of corrective control but also for the societal costs arising from the corrective control potential behavior modes.

To that end, we introduce a decision making approach on the basis of the socio-economic cost induced by a security management strategy. The socio-economic cost of a security management strategy comprises both of operational costs incurred by the TSO as well as of societal costs associated with the potential service disruption to the system users. In order to encapsulate the conceptual difference between the former and the latter, we consider the avoidance of potentially severe societal costs as an integral task of power system security management. We thus frame a probabilistic security management
concerning both economy and risk \cite{McCalley2004}, through  the following interrelated objectives:

\begin{itemize}
\item{\textbf{Primary Objective:} Avoid with a certain confidence the potential realization of extremely severe societal cost levels.}
\vspace{+0.1in}
\item{\textbf{Secondary Objective:} Minimize the expectation of the socio-economic costs of security under all credible system states.}
\end{itemize}

\subsection{Paper Organization}

We begin by presenting in section II the general form of the problem in question and discussing the main modeling issues that should be addressed. In the final part of this section, we also  introduce a set of approximations adopted to develop a tractable  algorithmic approach for the purposes of this paper. Based on such approximations, in section III we establish the detailed mathematical formulation of the considered problem as a Mixed-Integer Linear Programming (MILP) problem. In section IV, we demonstrate the properties of this proposal through a single-area academic test system and assess its effects on a comparative basis, with respect to the well-known N-1 approach. In section V we consider an alternative set of case studies concerning a system with two control areas. We conclude in section VI by summarizing the key findings of this work and discussing in detail the main challenges toward the adoption of probablistic power system security management practices.

\section{Security management in real-time operation}

In the present section we establish the general form of the proposed probabilistic real-time security management approach. Furthermore, we highlight the key modeling challenges and introduce a set of approximations employed in this paper for the sake of tractability.\footnote{For notational simplicity, but without loss in generality, we assume that all uncertainties range over a  finite number of possibilities.}

\begin{figure}
\centering
\includegraphics[width=0.49\textwidth]{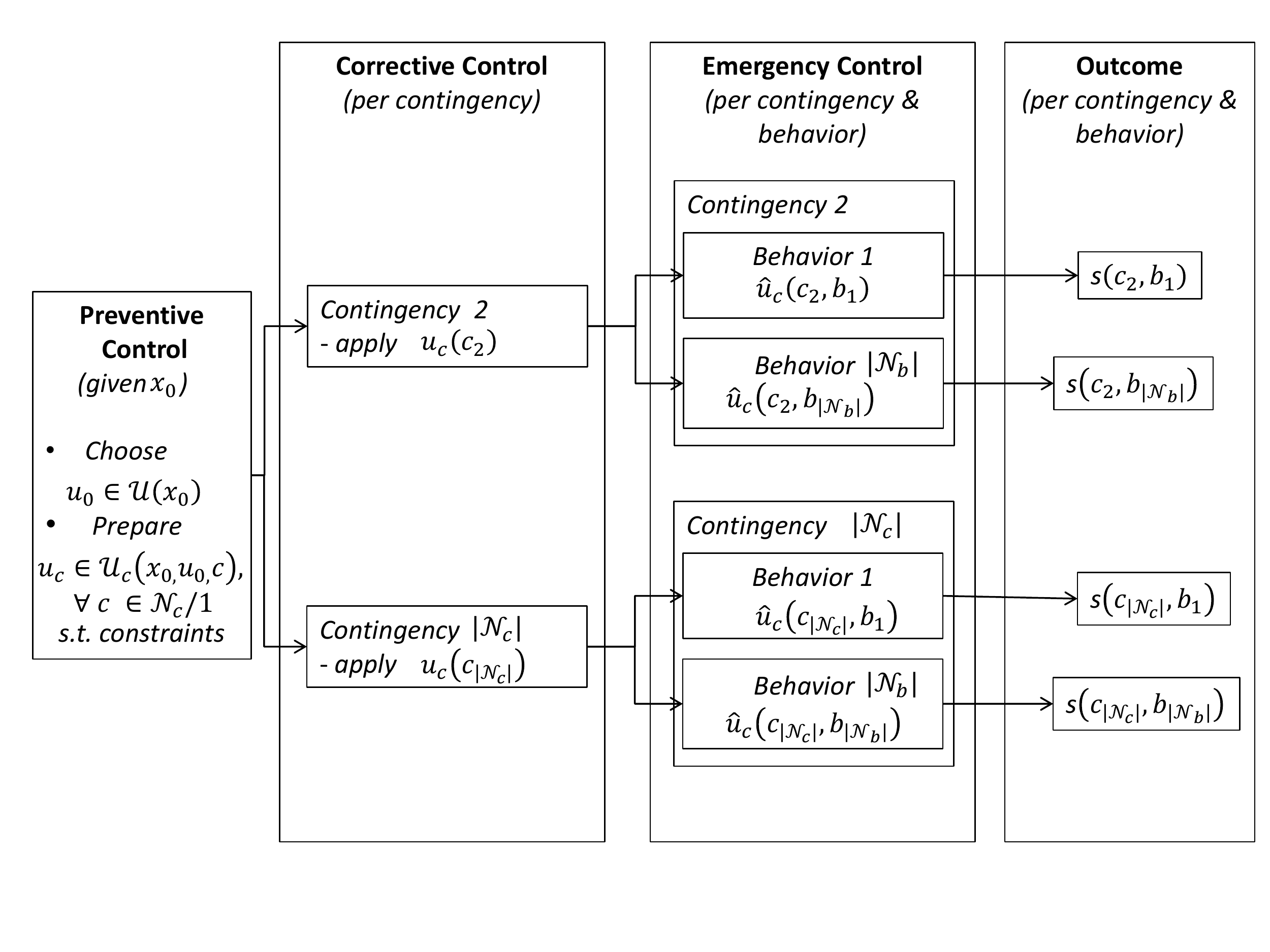}\vspace*{-8mm}
\caption{Decision making stages and outcomes}
\label{fig:framework}
\end{figure}

\subsection{Modelling the set of possible scenarios}

Let us denote by $\mathcal{N}_{c}$ the set of possible contingencies, by $x_{0}$ the current operating state of an interconnected system, by $\mathcal{U}_{0}(x_{0})$ the space of possible preventive control actions accessible to a TSO operating one area of this system, and for any pair $(u_{0}, c) \in \mathcal{U}_{0}(x_{0})\times \mathcal{N}_{c}$ by $\mathcal{U}_{c}(x_{0},u_{0},c)$ the space of possible corrective control actions accessible to {it}. (Please, preview Figure \ref{fig:framework}, before reading this section.)

We assume that the set of contingencies  $\mathcal{N}_{c}$ describes in a  mutually exclusive and exhaustive way the events that may happen during the next period of time $T$, and we denote by $\pi_{c} \in [0; 1]$ their elementary probabilities, with $\sum_{c\in \mathcal{N}_{c}} \pi_{c} = 1$.\footnote{The set $\mathcal{N}_{c}$ will thus in general contain also a pseudo-contingency which corresponds to the case where no actual contingency is triggered during the interval $T$.}

The job of the TSO is to choose a joint decision strategy $u$ combining a preventive control choice $u_{0} \in \mathcal{U}_{0}(x_{0})$ and a set of corrective control choices $\{u_{c}   \in \mathcal{U}_{c}(x_{0},u_{0},c)\}_{c \in \mathcal{N}_{c}}$. We denote by $\mathcal{U}(x_{0})$ the space of all joint preventive-corrective control strategies accessible to the TSO.
Once a joint strategy $u \in \mathcal{U}(x_{0})$ is chosen, the TSO applies its  {preventive decision} $u_{0}$ and then waits {for the potential realization of a contingency $c$ to apply  the corresponding corrective  control $u_{c}$. }

{Since corrective control is carried out under time pressure, and since the system state may change in an unpredictable way between the moment when a control action has been selected and the moment it is applied,} we do not assume that the physical effect of the control strategy is fully predictable \cite{fang2007}. Rather, we assume that for each four-tuple  $(x_{0}, u_{0}, c, u_{c})$ there is a set of possible post-contingency behaviors $\mathcal{N}_{b}(x_{0}, u_{0}, c, u_{c})$ that may occur; for each possible behavior $b \in \mathcal{N}_{b}(x_{0}, u_{0}, c, u_{c})$ we denote by $\pi_{b}(b|x_{0}, u_{0}, c, u_{c}) \in [0;1]$ its probability of occurrence, with $\sum_{b \in \mathcal{N}_{b}(x_{0}, u_{0}, c, u_{c})}\pi_{b}(b|x_{0}, u_{0}, c, u_{c}) = 1$.\footnote{In the most simple case, the set $\mathcal{N}_{b}$ would be composed of only two elements, one modelling perfect operation of corrective control, and one modelling complete failure of corrective control.}

\subsection{Security level induced by a decision strategy}

{We assume that the TSO can evaluate the impact of its decision strategies on power system security by computing a {\em severity function}  $s(x_{0}, u_{0}, c, u_{c}, b) \in \mathbb{R}^{+}$. Such a severity function would be computed} from the system dynamics induced by the choice of the preventive control action $u_{0}$, the occurrence of the contingency $c$, the application of the corrective control action $u_{c}$, and the realization of the post-contingency behavior $b$. This function would serve as a measure of the societal cost of the service disruptions realized by the end-users of {its} own control area.\footnote{{We note} that the computation of such a severity function {may} require very detailed dynamic simulations to identify the joint physical effect of the TSO's actions and post-contingency behaviors. Moreover, in order to express the physical quantities characterizing the resulting terminal state of the system in monetary terms,  various assumptions about the durations and costs of service interruptions would be required.}

For a given $x_{0}$ and a fixed strategy $u$, $s$ is thus a random variable. Let us define the two following quantities to describe its dependence on the choice of $u$:\footnote{{The operator $1(s \leq s')$ takes a value of one if $s\leq s'$, and otherwise takes a value of zero.}}
\begin{equation}
\hspace*{-2mm}\mathbb{P}_{s \leq s' | x_{0}, u}(u)  \hspace*{-1mm}=  \hspace*{-1mm}\sum_{c\in \mathcal{N}_{c}}\pi_{c}\hspace*{-2mm}\sum_{b\in\mathcal{N}_{b}}\hspace*{-2mm} \pi_{b}(b|x_{0}, u_{0}, c, u_{c}) 1(s \leq s'),
\end{equation}
which is the probability that the severity level $s$ is smaller or equal than the threshold $s'$, given the strategy $u$,
and
\begin{equation}
\mathbb{E}_{s | x_{0}, u} (u)  =  \sum_{c\in \mathcal{N}_{c}}\pi_{c}\sum_{b\in\mathcal{N}_{b}} \pi_{b}(b|x_{0}, u_{0}, c, u_{c}) s,
\end{equation}
which is the expected severity level induced by $u$.

\subsection{Formulating the optimal decision making problem}

As already stated, we propose to formulate the task of the {TSO} in real-time operation {security management} through a combination of two objectives:
\begin{itemize}
\item {\bf Primary objective:} avoid service interruptions of large severity to a certain possible extent. We model this objective by the choice of\textcolor{blue}{:} i) a maximal severity  level $s_{\max}$ to avoid and ii) a small risk $\epsilon$ of not being able to avoid it, i.e. in the form of the chance constraint:
\begin{equation}
\mathbb{P}_{s \leq s_{\max} | x_{0}, u}(u) \geq 1 - \epsilon . \label{cc}
\end{equation}
\item {\bf Secondary objective:} minimise a cost-function which combines the direct operating costs incurred by the  strategy and the societal costs measured by the expectation of the severity function:
\begin{eqnarray}
C(x_{0}, u) & = &  C_{0}(x_{0},u_{0}) +  \sum_{c\in \mathcal{N}_{c}}\pi_{c} C_{c}(x_{0}, u_{0}, c, u_{c}) \nonumber \\
& & +  \mathbb{E}_{s | x_{0}, u} (u) . \label{obj}
\end{eqnarray}
\end{itemize}
The real-time decision making problem ($\mbox{\bf RTP}$) hence is compactly expressed in the following way:\footnote{In this formulation all typical OPF constraints, such as static load-flow equations, control feasibility ranges, etc., are incorporated in the description of the set  of admissible joint  strategies $\mathcal{U}(x_{0})$.}\vspace*{2mm}

\centerline{\shadowbox{\parbox{8.5cm}{\begin{eqnarray}
\mbox{\bf RTP}: &\hspace*{-4mm} & \mbox{Compute:~} u^{*}(x_{0})  \in  \arg\min_{u} C(x_{0}, u) \\
&\hspace*{-4mm} &  \mbox{subject to:}  \left\{\begin{array}{l} u \in \mathcal{U}(x_{0}), \\  \mathbb{P}_{s \leq s_{\max} | x_{0}, u}(u) \geq 1 - \epsilon . \end{array}\right.
\end{eqnarray}}}}

\subsection{Scalable approximation strategies}

In large-scale power systems, the quantities $C(x_{0},u)$ and  $\mathbb{P}_{s \leq s_{\max} | x_{0}, u}(u)$ used in our formulation can't be calculated exactly, even for a single ``given'' strategy $u$ (and even if we consider that $\mathcal{N}_{c}$ and $\mathcal{N}_{b}$ are finite sets). Hence the exact solution of the $\mbox{\bf RTP}$ optimisation problem is certainly out of reach in realistic conditions.

As a starting point in the direction of providing tractable solutions to this problem, in the present paper we adopt the following approximations:

\begin{itemize}
\item[(i)]{We express all network constraints under the DC power flow model \cite{dcflow}. On this basis, we consider the pre-contingency scheduling of the generating units as the preventive control action available to the TSO. Following the occurrence of any contingency $c\ge 1$, the set of available control actions corresponds to the re-dispatch of the generating units that were operational in the pre-contingency state. }
\item[(ii)]{We restrict to considering that the set $\mathcal{N}_b$ comprises solely of two elements, corresponding to the working and failing behaviors of corrective control respectively. Under the latter, corrective control would be completely ineffectual. In such a case, all controllable resources (\textit{i.e.} dispatchable generating units) would remain at their preventive operating points.}
\item[(iii)]{Following from approximation (ii), the realization of the corrective control failing behavior under the case of a generating unit outage would result in a net energy deficit.   In such a case, we seek for a terminal system state, as in approximation (v), by considering that the transmission network remains intact.}

\item[(iv)]{Also following from approximation (ii), the realization of the corrective control failing behavior under the case of a line outage may result in overloading additional network branches. In such a case, prior to identifying the terminal system state we model the removal of such branches from service.}

\item[(v)]{Under both line and generating unit outages, we seek for a feasible terminal system state following the realization of corrective control behavior $b \in \mathcal{N}_b$ by means of the following emergency control actions:
\begin{itemize}
\item{Shedding load across any network node.}
\item{Allowing generating units to ramp-down within a limited range in response to the load shedding. We note that in the case that the corrective control failing behavior was realized, this limited range is applicable with respect to the preventive operating points of all generating units that were not affected by the initiating contingency.}
\item{Imposing that in case a generating unit cannot operate within the aforementioned limited range it would be disconnected from the network.}
\end{itemize}}
\item[(vi)]{Following the application of the emergency control actions listed in approximation (v) we adopt a linear severity function to express the societal cost of the service disruption. In this function we employ a load-specific per unit free (in \euro/$MWh$) to denote the cost incurred by any electricity consumer subject to a service interruption. With regard to generating units, we do not consider any emergency control cost in the event that they are merely forced to ramp-down within the aforementioned limited range. We recall that function \eqref{obj} explicitly accounts for the costs of the preventive and corrective dispatch under every credible system state. As such, we restrict to considering a unit specific fixed fee (in \euro) in the event that a generating unit has to be disconnected in order to reach a feasible terminal state. Such fee would reflect not only the shut-down cost of a generating unit but also the fact that the unexpected disconnection may disrupt the unit's planned operating schedule for the subsequent periods due to minimum down-time restrictions. Evidently, the level of this fee would depend both on the technical characteristics of every generating unit (\textit{i.e.} capacity, flexibility \textit{etc.}) as well as on its connection point to the network.}
\item[(vii)]{Given that through \eqref{cc} and \eqref{obj} the present formulation explicitly controls the severity arising from the realization of any contingency $c \in \mathcal{N}_c$ and every possible post-contingency corrective control behavior $b \in \mathcal{N}_b$, we allow for the relaxation of the post-contingency transmission capacity limits under the corrective control working behavior. In the event that such limits have been relaxed at the optimal solution, we model the consequences of such a decision as described in approximation (iv), \textit{i.e.} we remove the overloaded branches from service prior to seeking for a feasible terminal state. }
\end{itemize}

The effect of all such approximations on the findings of this paper, as well as the potential for reconsideration will be discussed in detail at the final parts of this paper.

\section{Mathematical Formulation}
\label{section:math}

In the present section we introduce the detailed form of the security management problem under consideration. We note that problem \eqref{of}-\eqref{sev} can be cast as a Mixed-Integer Linear Programming Problem (MILP) by employing the linearization technique detailed in appendix A.
\begin{flalign}
\min &\left\{\sum_{g \in N_g}c_g \cdot P_{g}^0+ \sum_{c \in N_c \setminus 1}\pi_{c} \cdot c^r_g \left(P_{g}^c-P_g^0\right)
\right. & \nonumber \\
&\quad \left.+\sum_{c \in N_c \setminus 1} \pi_{c} \cdot \sum_{b \in N_{b}} \pi_{b} \cdot s^c(b)\right\}  \label{of}&
\intertext{subject to,}
\intertext{$\quad$ for all nodes $n \in \mathcal{N}_n$:}
&\sum_{g \in \mathcal{G}_{n}}P_{g}^0-\sum_{{\ell} \in \mathcal{N}_{\ell}}\beta_{n,\ell} \cdot f_{{\ell}}^0=\sum_{d \in \mathcal{D}_{n}}P_{d}^0& \label{pre_pbal}\\
\intertext{$\quad$for all lines $\ell \in \mathcal{N}_{\ell}$}
&f_{\ell}^0- \frac{1}{X_{\ell}}\sum_{n \in \mathcal{N}_n} \beta_{n,\ell} \cdot \theta_{n}^0 =0&\label{pre_pf}\\
&f_{\ell}^0\leq f_{\ell}^{max}&  \label{pre_posflow}\\
&-f_{\ell}^0\leq f_{\ell}^{max}&   \label{pre_negflow}\\
\intertext{$\quad$for all generators $g \in \mathcal{N}_g$}
&P_{g}^0\leq P_{g}^{max}&  \label{pre_uppgen}\\
&-P_{g}^0\leq -P_{g}^{min}&  \label{pre_lowgen}\\
\intertext{$\quad$for all nodes $n \in \mathcal{N}_n$,  all contingencies $c\in\mathcal{N}_c$ and corrective control working behavior:}
&\sum_{g \in \mathcal{G}_{n}}P_{g}^c-\sum_{{\ell} \in \mathcal{N}_{\ell}}\beta_{n,\ell} \cdot f_{{\ell}}^c(b)=\sum_{d \in \mathcal{D}_{n}}P_{d}^0& \label{post_pbal_one}\\
\intertext{$\quad$for all nodes $n \in \mathcal{N}_n$, all contingencies $c\in\mathcal{N}_c$ and corrective control failing behavior:}
&\sum_{g \in \mathcal{G}_{n}}a_g^c \cdot P_{g}^0-\sum_{{\ell} \in \mathcal{N}_{\ell}}\beta_{n,\ell} \cdot f_{{\ell}}^c(b)+\delta_{n}^{c}(b)=\sum_{d \in \mathcal{D}_{n}}P_{d}^0& \label{post_pbal_two}\\
&\delta_{n}^{c}(b)- \tau_c \cdot \sum_{g \in \mathcal{G}_{n}}(P_g^c-P_{g}^0)=0&\label{delta}\\
\intertext{$\quad$for all lines $\ell \in \mathcal{N}_{\ell}$, all contingencies $c\in\mathcal{N}_c$ and all corrective control behaviors $b \in N_b$:}
&f_{\ell}^c(b)-\frac{a_{\ell}^c}{X_{\ell}}\sum_{n \in \mathcal{N}_n} \beta_{n,\ell} \cdot \theta_{n}^c(b) =0 &\label{relax_pf}\\
&f_{\ell}^c(b)-a_{\ell}^c \cdot \lambda_{\ell}^c(b) \cdot M \leq a_{\ell}^c \cdot f_{\ell}^{max}    & \label{relax_posflow}\\
&-  f_{\ell}^c(b)-a_{\ell}^c \cdot \lambda_{\ell}^c(b) \cdot M  \leq a_{\ell}^c \cdot f_{\ell}^{max}    &\label{relax_negflow}\\
&f_{\ell}^c(b)- p_{\ell}^c(b) \cdot M\leq 0&\label{flow_sign_one}\\
& -f_{\ell}^c(b)-\left(1- p_{\ell}^c(b)\right) \cdot M\leq0 &\label{flow_sign_two}\\
&\lambda_{\ell}^c(b) -\frac{f_{\ell}^c(b)}{f_{\ell}^{max}+\varepsilon}-\left(1- p_{\ell}^c(b)\right) \cdot M\leq0 &\label{up_lam_one}\\
&\lambda_{\ell}^c(b) +\frac{f_{\ell}^c(b)}{f_{\ell}^{max}+\varepsilon}-p_{\ell}^c(b) \cdot M\leq0 &\label{up_lam_two}\\
\intertext{$\quad$for all generators $g \in \mathcal{N}_g$ and all contingencies $c\in \mathcal{N}_c$ }
&P_{g}^c \leq a_{g}^c \cdot   P_{g}^{max}   &\label{post_uppgen}\\
&-P_{g}^c \leq - a_{g}^c \cdot  P_{g}^{min}   &\label{post_lowgen}\\
&P_{g}^{c}-a_{g}^c \cdot P_{g}^0  \leq a_{g}^c \cdot  P_{g}^{+} &\label{post_uppcoup}\\
&a_{g}^c \cdot P_{g}^0-P_{g}^{c}  \leq  a_{g}^c \cdot  P_{g}^{-} & \label{post_dwncoup}\\
\intertext{$\quad$for all nodes $n \in\mathcal{N}_n$, all contingencies $c\in\mathcal{N}_c$ and all corrective control behaviors $b \in N_b$:}
&\sum_{g \in \mathcal{G}_{n}}\hat{P}_{g}^c(b)-\sum_{{\ell} \in \mathcal{N}_{\ell}}\beta_{n,\ell}\cdot \hat{f}_{\ell}^c(b)-\sum_{d \in \mathcal{D}_{n}}\hat{P}_{d}^c(b) =0&\label{low_pbal}\\
\intertext{$\quad$for all lines $\ell \in \mathcal{N}_{\ell}$,  all contingencies $c\in\mathcal{N}_c$ and  corrective control working behavior:}
&\hat{f}_{\ell}^c(b)- \frac{a_{\ell}^c \cdot (1-\lambda_{\ell}^c(b)) }{X_{\ell}}\sum_{n \in \mathcal{N}_n} \beta_{n,\ell} \cdot \hat{\theta}_{n}^c(b)  =0&\label{low_pf}\\
&\hat{f}_{\ell}^c(b)- a_{\ell}^c \cdot (1-\lambda_{\ell}^c(b)) \cdot f_{\ell}^{max}\leq 0&  \label{low_posflow}\\
&-\hat{f}_{\ell}^c(b)- a_{\ell}^c \cdot (1-\lambda_{\ell}^c(b)) \cdot  f_{\ell}^{max} \leq 0&   \label{low_negflow}\\
\intertext{$\quad$for all lines $\ell \in \mathcal{N}_{\ell}$, all contingencies $c\in\mathcal{N}_c$ and  corrective control failing behavior:}
&\hat{f}_{\ell}^c(b)- \frac{a_{\ell}^c \cdot (1-\tau_c\cdot \lambda_{\ell}^c(b)) }{X_{\ell}}\sum_{n \in \mathcal{N}_n} \beta_{n,\ell} \cdot \hat{\theta}_{n}^c(b)  =0&\label{low_pf2}\\
&\hat{f}_{\ell}^c(b) - a_{\ell}^c \cdot (1-\tau_c\cdot \lambda_{\ell}^c(b)) \cdot f_{\ell}^{max} \leq 0&  \label{low_posflow2}\\
&-\hat{f}_{\ell}^c(b) -  a_{\ell}^c \cdot (1-\tau_c\cdot \lambda_{\ell}^c(b)) \cdot  f_{\ell}^{max} \leq 0 &   \label{low_negflow2}\\
\intertext{$\quad$for all demands $d \in \mathcal{N}_n$, all contingencies $c\in\mathcal{N}_c$ and corrective control behaviors $b \in N_b$:}
&\hat{P}_{d}^c(b)\leq  P_{d}^{0}     &\label{low_dem}
\intertext{$\quad$for all generators $g \in \mathcal{N}_g$, all contingencies $c\in\mathcal{N}_c$ and corrective control working behavior:}
&\hat{P}_{g}^c(b) -  \left(1-y_g^c(b)\right) \cdot  P_g^c  \le 0\label{low_uppcoup}\\
&-\hat{P}_{g}^c(b)+\left(1-y_g^c(b)\right)  \cdot\left( P_g^c  -a_{g}^c \cdot  \Delta P_{g}^{e}\right)  \le 0 \label{low_dwncoup}\\
\intertext{$\quad$for all generators $g \in \mathcal{N}_g$, all contingencies $c\in\mathcal{N}_c$ and corrective control failing behavior:}
&\hat{P}_{g}^c(b) -  a_g^c \cdot \left(1-y_g^c(b)\right) \cdot  P_g^0  \le 0\label{low_uppcoup2}\\
&-\hat{P}_{g}^c(b)+a_g^c \cdot \left(1-y_g^c(b)\right) \cdot \left( P_g^0  -  \Delta P_{g}^{e}\right)  \le 0 \label{low_dwncoup2}\\
\intertext{$\quad$for all generators $g \in \mathcal{N}_g$, all contingencies $c\in\mathcal{N}_c$ and corrective control behaviors $ b \in \mathcal{N}_b$ :}
&-\hat{P}_{g}^c(b) + a_g^c \cdot \left(1-y_g^c(b)\right) \cdot  P_g^{min}  \le 0\label{low_mingen}\\
\intertext{$\quad$ for all contingencies $c\in\mathcal{N}_c$ and corrective control behaviors $ b \in \mathcal{N}_b$ :}
&-s^c(b) + \sum_{d \in \mathcal{N}_d}v_d\cdot (P_d^0-\hat{P}_{d}^c(b))+\sum_{g \in \mathcal{N}_g}w_g\cdot y_g^c(b) \le 0& \label{sev}\\
&s^c(b) -  \gamma^c(b) \cdot M   \le s_{\max}& \label{chance_one} \\
&\sum_{c \in \mathcal{N}_c}\pi_c \cdot \sum_{b \in \mathcal{N}_{b}} \pi_{b} \cdot \gamma^c(b)  \le \epsilon&\label{chance_two}
\end{flalign}

The first term in \eqref{of} expresses the cost of the preventive generation schedule while the second denotes the expected cost of the corrective re-dispatch. The third term is the expectation of the severity function. We underline that the form of the first two terms of \eqref{of} depends on the applicable regulatory framework. Without loss in generality, through function \eqref{of} we consider the case wherein a TSO seeks to minimize the expected net generation cost under any credible event. The extension of this work to alternative regulatory arrangements, \textit{e.g.} in the case wherein the TSO is accountable for the cost of changes with respect to the outcome of an already settled market clearing, remains straightforward (see appendix B). 

 The pre-contingency nodal power balance is enforced via equality \eqref{pre_pbal}. Expressions \eqref{pre_pf} to \eqref{pre_negflow} correspond to the DC power flow approximation and line capacity limits in the pre-contingency state respectively. Moreover, the feasible operating region of the generating units in the pre-contingency state is expressed by  \eqref{pre_uppgen}-\eqref{pre_lowgen}.

 The post-contingency nodal power balance is expressed through \eqref{post_pbal_one} and \eqref{post_pbal_two}. The former holds for the case that there is no failure in the corrective control, hence the equality is strictly enforced. We underline that,  in case  the contingency relates to the outage of a generating unit, the ineffectiveness of the corrective re-dispatch would result in an energy mismatch. To account for this effect in \eqref{post_pbal_two}, we introduce a free slack variable $(\delta_n^c(b))$  per node. Equality \eqref{delta} defines the value of such slack variable with the use of auxiliary binary parameter $\tau_c$, which would only take a value of 1 for a generator outage.

The DC power flow approximation in the post-contingency states is expressed by \eqref{relax_pf}. Binary variable $\lambda_{\ell}^c(b)$ is used in \eqref{relax_posflow} and \eqref{relax_negflow} to allow for the relaxation of the line capacity limits. To achieve that this variable would only take a value of one when the capacity limits are exceeded, logical constraints \eqref{flow_sign_one} through \eqref{up_lam_two} are used\footnote{Symbol $\varepsilon$ denotes an infinitesimally small constant.}. More specifically, \eqref{flow_sign_one} - \eqref{flow_sign_two} enforce that the value of binary variable $p_{\ell}^c(b)$  would be equal to one if the flow is positive (zero if the flow is negative). In the former case, \eqref{up_lam_one} imposes that binary variable $\lambda_{\ell}^c(b)$  can only be equal to one if the flow is (at least infinitesimally) greater than the corresponding limit, while \eqref{up_lam_two}  is inactive. Conversely, in the case that the sign of $f_{\ell}^c(b)$ is negative, \eqref{up_lam_two}  takes force to ensure that a non-zero value of binary variable $\lambda_{\ell}^c(b)$  indicates an overload while \eqref{up_lam_one} is inactive.

The feasible operating region of the generating units in the post- contingency states is expressed via \eqref{post_uppgen}-\eqref{post_dwncoup}. Coupling constraints \eqref{post_uppcoup} and \eqref{post_dwncoup} are used to consider ramping restrictions.

In order to quantify the severity of the system state, we seek for a feasible power flow solution following the realization of the effects of corrective control behavior. More specifically, expression \eqref{low_pbal} re-establishes the nodal power balance constraint. Furthermore, \eqref{low_pf}-\eqref{low_negflow2} impose that overloaded network branches would be disconnected\footnote{We should clarify that auxiliary binary parameter $\tau_c$ is used in \eqref{low_pf2}-\eqref{low_negflow2} to maintain all network branches in service in the case that corrective control failed following a generator outage. This is due to the fact that, as explained earlier, in such a case the nodal power balance in the post-contingency state \eqref{post_pbal_two} can only be achieved by introducing a set of slack variables.}. We note that the  products of binary and continuous variables appearing in \eqref{low_pf}-\eqref{low_negflow2} can be linearized according to the methodology presented in appendix A.

The set of admissible emergency control actions is modeled via \eqref{low_dem}-\eqref{low_mingen}. Inequality \eqref{low_dem} allows the use of load shedding as a measure of last resort. Constraints \eqref{low_uppcoup} through \eqref{low_dwncoup2}   impose that the generating units should either only ramp-down within a limited range (following the shedding of load) or be disconnected from the network.  More specifically, \eqref{low_uppcoup} and \eqref{low_dwncoup} apply in the case wherein corrective control has been effective. In such a case, the emergency ramp-down limit would apply with respect to the post-contingency dispatch of any generating unit. Alternatively, in the case that corrective control has proven ineffectual, the limit is applicable with respect to any unit's preventive dispatch, as in \eqref{low_uppcoup2} and \eqref{low_dwncoup2}. In both cases, if a generating unit has to ramp down beyond the admissible limit, binary variable $y_g^c(b)$ would take a value of one to denote the disconnection of this unit\footnote{The linearization of products of binary and continuous variables in \eqref{low_uppcoup} and \eqref{low_dwncoup}  is presented in Appendix A as well.}. Likewise, \eqref{low_mingen} expresses the minimum stable generation restriction which remains valid.

Taking these measures used to reach a terminal state into account, the severity of the system state is expressed as in \eqref{sev}. The first term of \eqref{sev} gathers the value of the load that cannot be served. As already mentioned, through the second term, we consider a fixed cost in the case that a generating unit has to be disconnected from the network. We recall that, parameter $w_g$ essentially denotes the importance of ensuring the continuity in the connection of unit $g$ to the network. Such parameter would be unit specific and depend not only on the capacity of the unit but also on its location in the network and its flexibility characteristics. Finally, inequalities \eqref{chance_one} and \eqref{chance_two} impose the chance constraint on the severity level through the use of auxiliary binary variable $\gamma^c(b)$.

\section{Single-Area Case Study}

\begin{figure}
\centering
\includegraphics[width=0.38\textwidth]{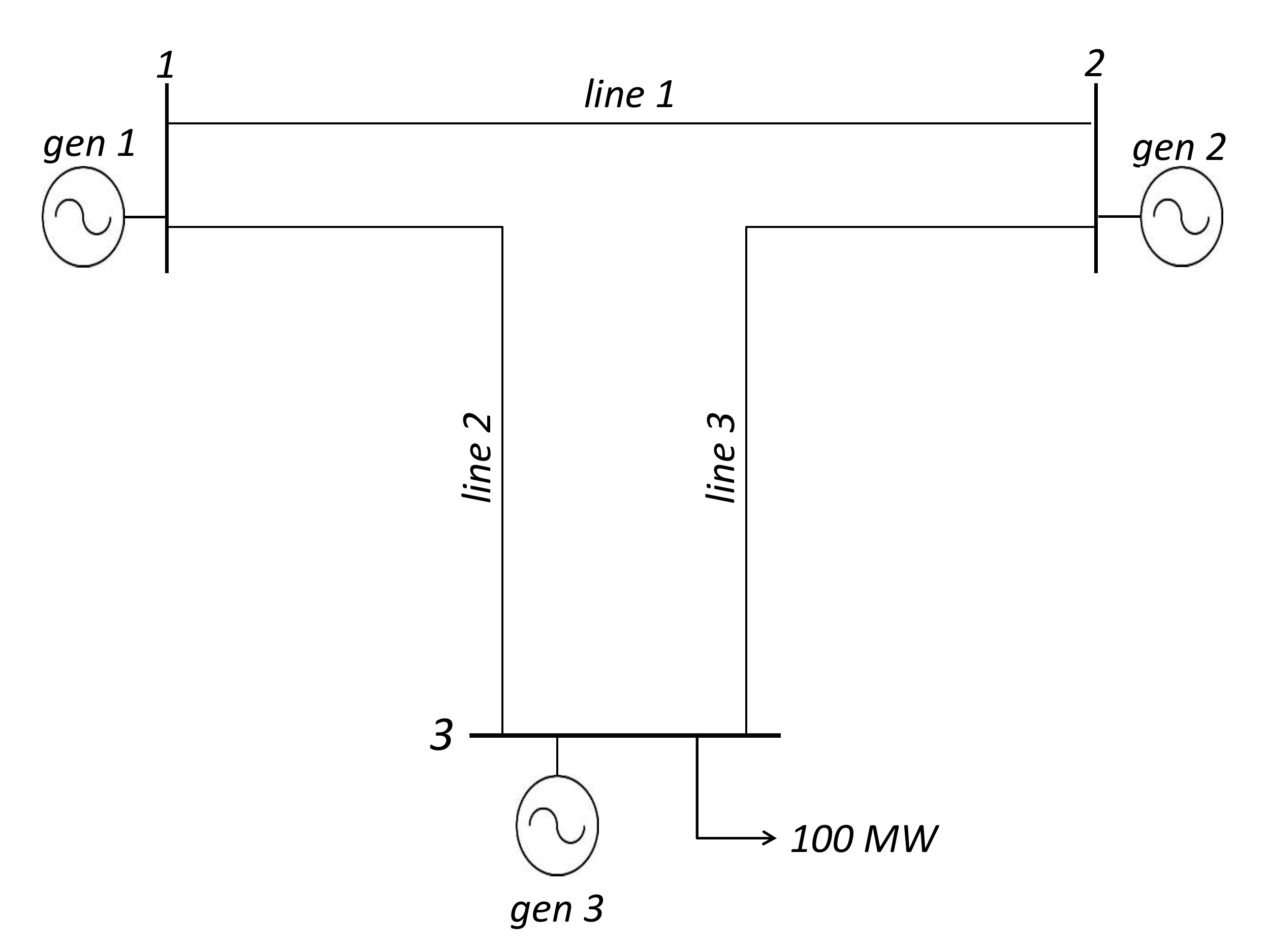}
\caption{Three-node, three-generator system}
\label{fig:3bus}
\end{figure}

In the interest of establishing the novel features of the proposed approach we begin by discussing a set of demonstrative case studies based on the 3-bus network of figure \ref{fig:3bus}. Table \ref{table:gens} summarizes the parameters of the 3 generating units\footnote{For the sake of simplicity, ramping restrictions have been omitted in this example.}. All network branches are assumed to have a maximum capacity of $55$ MW, equal admittance and a mean time to failure of 10000 h, as in \cite{bouff2}. The demand at node 3 is assumed to be equal to 100 MW for a period of 1 hour, while the value of lost load is set at \euro 300/MWh.

We underline that for the purposes of these demonstrative cases studies and without loss in generality, both the value of lost load $(v_d)$ as well as the generating units' disconnection severity coefficients ($w_g$) have been arbitrarily set. In the interest of signifying the value of maintaining service to the network users, the former was set  an order of magnitude greater than the per unit generation costs in the test system. Recalling that the latter should reflect costs incurred at the period of disconnection as well as the criticality of disrupting the availability of supply for the forthcoming periods, a considerably large value was selected. The critical role of such parameters on the performance of the proposed approach, as well as the necessity of establishing credible reference values, will be discussed in detail at the concluding parts of this paper.

\begin{table}
\caption{Generating Units Specifications }
\centering
\def~{\phantom{0}}
\begin{tabular}{ccccccc}
\sphline
\multirow{2}{*}{$g$}& $c_{g}$ & $c_{g}^r$ & $P_g^{min}$& $P_g^{max}$&MTTF&$w_g$\\
& \multicolumn{2}{c}{(\euro/$MWh$)}&\multicolumn{2}{c}{( $MW$)}&($h$)&(\euro)\\
\sphline
1 & 20 & 5 & 10&100&500&4000\\
2 & 40 & 8 & 10&100&500&4000\\
3 & 30 & 7 & 10&50&250&4000\\
\sphline
\end{tabular}
\label{table:gens}
\end{table}

The set of credible failures considered in this example involves the failure of any single component. The probabilities of all credible contingencies, including the pseudo-contingency of no failure, have been calculated according to \cite{bilir} and are listed in table \ref{table:probs}. Following the occurrence of any contigency, we assume that the probability of realization of the corrective control working behavior is 0.8.

\begin{table}
\caption{Contingency Set }
\centering
\def~{\phantom{0}}
\begin{tabular}{ccc}
\sphline
c& Event & $\pi_c$\\
\sphline
1 & No Outage& 0.99193\\
2 & Line 1 Outage& $0.9\cdot 10^{-4}$\\
3 & Line 2 Outage& $0.9\cdot 10^{-4}$\\
4 & Line 3 Outage& $0.9\cdot 10^{-4}$\\
5 & Gen. 1 Outage& $1.9\cdot 10^{-3}$\\
6 & Gen. 2 Outage& $1.9\cdot 10^{-3}$\\
7 & Gen. 3 Outage& $4\cdot 10^{-3}$\\
\sphline
\end{tabular}
\label{table:probs}
\end{table}

\subsection{Case A: The `hidden' severity levels of the N-1 approach}

In order to establish a benchmark for comparison, we begin by presenting the sequence of preventive and corrective actions under the  N-1 approach. This sequence has been identified by simplifying the formulation presented in section \ref{section:math} in the following ways:

\begin{itemize}
\item{The third term of \eqref{of} has been omitted.}
\item{A sufficiently large admissible severity threshold, making constraints \eqref{chance_one}-\eqref{chance_two} ineffectual, has been selected.}
\item{All constraints referring to the corrective control working behavior have been strictly enforced, while all constraints referring to the corrective control failing behavior can be relaxed.}
\end{itemize}

Owing to these simplifications, decision making in case A follows the assumption that corrective control is fully reliable. Nevertheless, the effects of the possible failure of corrective control, as well as the resulting severity levels are computed as by-products of the optimization process.

\begin{table}
\centering
\caption{Case A: Preventive Dispatch ($MW$) }
\label{caseAprev}
\begin{tabular}{ccccc}
\sphline
$g$&1&2&3\\
\sphline
&77.5&10&12.5\\
\sphline
\end{tabular}
\end{table}

\begin{table}
\centering
\caption{Case A: Corrective Re-dispatch ($MW$) }
\begin{tabular}{ccccc}
\sphline
$c\setminus g$&1&2&3\\
\sphline
2&55&10 & 35\\
3&45& 10& 45\\
4&45& 10& 45\\
5&x  & 50&  50\\
6&82.5& x &17.5\\
7&65& 35&   x\\
\sphline
\end{tabular}
\label{caseAcorr}
\end{table}

The preventive generation schedule as well as the corrective re-dispatch under any credible contingency are presented in tables \ref{caseAprev} and \ref{caseAcorr} respectively. Evidently, under the corrective control working behavior the system expected severity is zero since there exists a combination of preventive and corrective actions to serve the demand under any contingency. Nevertheless,  it is also evident that the sequence of actions listed in tables \ref{caseAprev} and \ref{caseAcorr} is not resilient against the possibility that the corrective re-dispatch may not materialize.

In fact, for generating unit outages the occurrence of an energy deficit would be inevitable. Acknowledging that all line outages ($c=2,\dots,4$) would require re-dispatch actions (by inspecting table \ref{caseAprev} and rows 1-3 of table \ref{caseAcorr}) we can identify that network capacity ratings would be exceeded in case corrective control failed to take effect. Indeed, this is confirmed in table \ref{caseAflow_fail}, wherein the line flow patterns under the corrective control failing behavior are presented for these contingencies.

\begin{table}
\centering
\caption{Case A: Power Flows under Corrective Control Failure ($MW$) }
\label{caseAflow_fail}
\begin{tabular}{ccccc}
\sphline
$c\setminus \ell$&1&2&3\\
\sphline
2&x&\textbf{77.5} & 10\\
3&\textbf{77.5} & x & \textbf{87.5} \\
4&-10& \textbf{87.5} & x\\
\sphline
\end{tabular}
\end{table}

\begin{figure}
\centering
\includegraphics[width=0.38\textwidth]{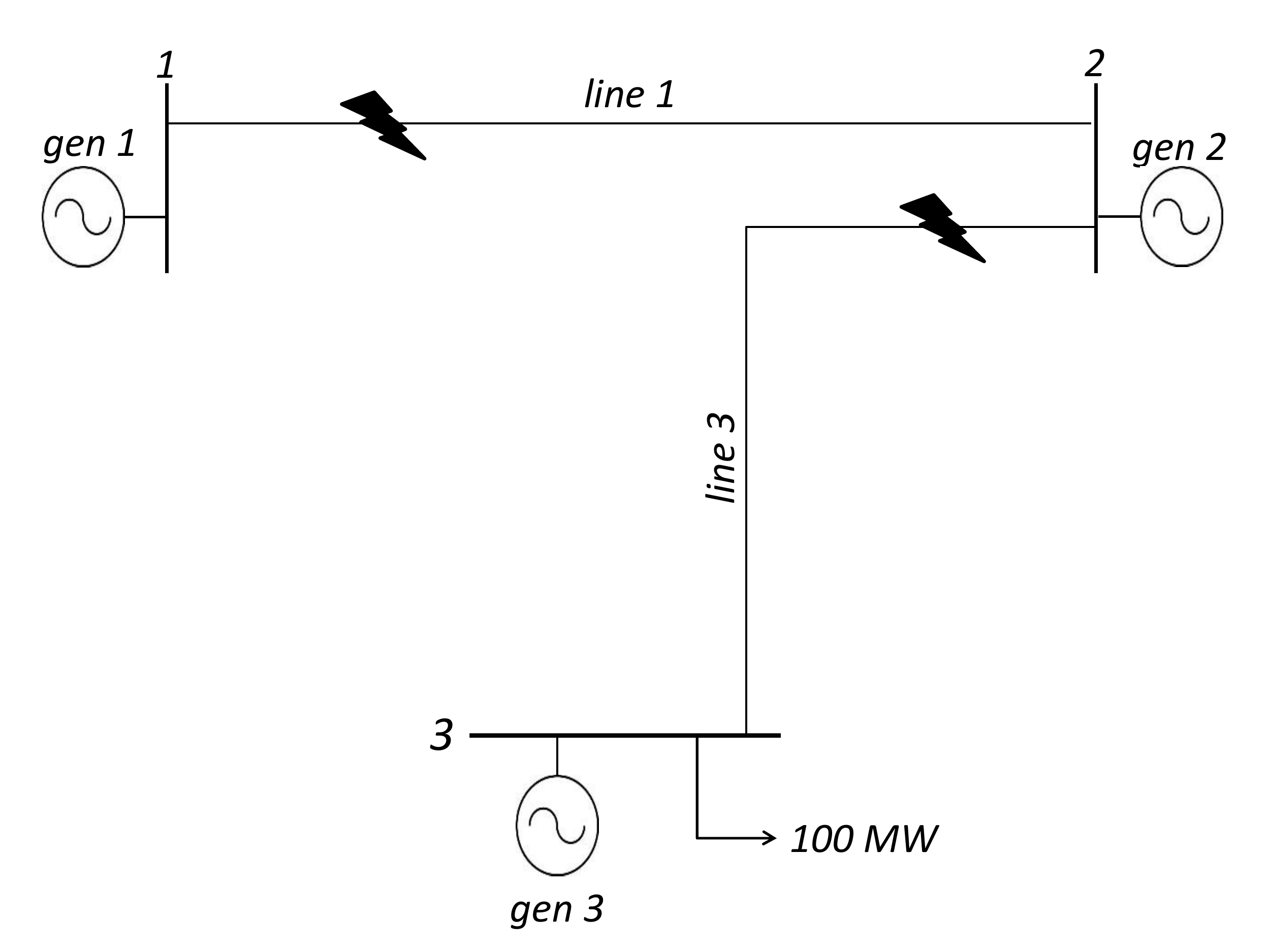}
\caption{Failure of Corrective Control under Outage of Line 2}
\label{fig:3bus_cng2}
\end{figure}

Table \ref{caseAsevl} presents the potential severity levels  (first row)  along with the corresponding probabilities of realization (third row). As an example, figure \ref{fig:3bus_cng2} illustrates the potential effect of corrective control failure following contingency 3. In such a case, the failure of corrective control would result in overloading both network branches that were not affected by the initiating contingency. Acknowledging that these overloaded branches would be subsequently taken out of service, all 3 network nodes would eventually be isolated. Consequently, generating units 1 and 2 would have to be disconnected due to minimum stable generation restrictions. Moreover, given that controllability of generating unit 3 is assumed to be lost, the maximum available generation to serve the demand is $12.5 MW$, \textit{i.e.} the pre-contingency output of unit 3. As a result $87.5 MW$ of demand would have to be shed for a period of 1 hour.

\begin{table}
\centering
\caption{Case A: Severity Levels}
\label{caseAsevl}
\begin{tabular}{ccccccc}
\sphline
$c$&2&3&4&5&6&7\\
\sphline
$s^c(b)$ (\euro)& 27250&34250&34250&23250&3000&3750\\
$s^c(b)$ (\%)&64.88&81.55&81.55&55.36&7.14&8.92 \\
$\pi_c \cdot \pi_b \cdot 10^{-4}$&0.18&0.18&0.18&3.8&3.8&8\\
\sphline
\end{tabular}
\end{table}

Noting that the maximum possible severity level in the present example is equal to \euro 42000 (in the case that all units are disconnected and none of the demand is served) the second row of table \ref{caseAsevl}  presents a relative comparison. We should conclude by stating that the cost of the preventive generation schedule in this case is \euro 2325 while the expected cost of the corrective redispatch is \euro 0.55. Relative to the latter figure, the expected severity level of \euro 14.7 appears non-negligible.

\subsection{Case B:  Acknowledging the Severity of Corrective Control Potential Failure }

The second case study serves to demonstrate the effect of considering the potential failure of corrective control via controlling the system severity levels. To that end, we now consider the complete formulation presented in section \ref{section:math}. For the sake of consistency, we begin by maintaining from case  A the constraint that the line capacity limits should not be relaxed under the corrective control working behavior. The potential relaxation of these capacity limits will be  investigated at a subsequent part of this case study.

The  admissible severity threshold is set at 33\% $(1/3)$ of the largest possible severity value, \textit{i.e.} $s_{max}=14000$ \euro. We begin by considering this threshold as a hard constraint by setting the upper bound on the severity violation probability to zero. The effect of chance constraint \eqref{chance_one}-\eqref{chance_two} on the severity threshold violation will also be demonstrated at a following part of this case study.

\begin{table}
\caption{Case B: Preventive Dispatch  ($MW$) }
\label{BcaseBprev}
\centering
\begin{tabular}{ccccc}
\sphline
$g$&1&2&3\\
\sphline
&45&10&45\\
\sphline
\end{tabular}
\vspace{+0.1 in}
\caption{Case B: Corrective Re-dispatch ($MW$) }
\label{BcaseBcorr}
\begin{tabular}{ccccc}
\sphline
$c\setminus g$&1&2&3\\
\sphline
2&55&10 & 35\\
3&45& 10& 45\\
4&45& 10& 45\\
5&x  & 50&  50\\
6&82.5& x &17.5\\
7&65& 35&   x\\
\sphline
\end{tabular}
\end{table}

Tables \ref{BcaseBprev} and \ref{BcaseBcorr} present the preventive dispatch and corrective re-dispatch. In comparison to case A (tables \ref{caseAprev} and \ref{caseAcorr}) we can identify that the corrective re-dispatch remains unaltered. This was anticipated given the fact that in case A, the severity levels under the corrective control working behavior were equal to zero  for all contingencies. Likewise as expected, the difference between cases A and B lies in the preventive schedule. The preventive schedule implemented in case B comes at an increased cost since generating unit 3 is now displacing cheaper unit 1.

Such cost increase comes at the benefit of ensuring that the severity level is always below the desirable threshold, even under the potential failure of corrective control. In order to achieve this requirement, the selected  preventive dispatch is feasible under all line outages, table \ref{caseBflow_fail}. Given this fact, it is noteworthy that a corrective re-dispatch has also been selected. This is due to the fact that in case the corrective control does not fail to have an effect, the re-dispatch would generate a cost reduction. In this sense, the secure preventive dispatch serves as a fall-back for the case that corrective control is ineffectual. As a result, the severity under corrective control failure for all line outages has been eliminated, table \ref{caseBsevl}. Nevertheless,  the energy mismatches arising from generation outages when corrective control is ineffectual can only be counteracted by means of emergency load shedding. As such, the severity levels for all generating unit outages remain non-zero. On average, these levels are lower than case A though.

\begin{table}
\centering
\caption{Case B: Power Flows under Corrective Control Failure ($MW$) }
\label{caseBflow_fail}
\begin{tabular}{ccccc}
\sphline
$c\setminus \ell$&1&2&3\\
\sphline
2&x&45 & 10\\
3&45 & x & 55 \\
4&-10& 55 & x\\
\sphline
\end{tabular}
\end{table}

\begin{table}
\centering
\caption{Case B: Severity Levels}
\label{caseBsevl}
\begin{tabular}{ccccccc}
\sphline
$c$&2&3&4&5&6&7\\
\sphline
$s^c(b)$ (\euro)& 0&0&0&13500&3000&13500\\
$s^c(b)$ (\%)&0&0&0&32.14&7.14&32.14 \\
$\pi_c \cdot \pi_b \cdot 10^{-4}$&0.18&0.18&0.18&3.8&3.8&8\\
\sphline
\end{tabular}
\end{table}

Finally, in table \ref{breakdown_costs} we compare the preventive and expected corrective costs, as well as the expected severity levels between cases A and B. As anticipated, the preventive cost of case B is greater than that of case A. It should be noted that this cost increase is asymmetrical to the reduction of the expected severity level. This asymmetry serves to demonstrate the conceptual difference between the monetary and non-monetary costs of power system security. We recall that in the present case, the admissible severity threshold was considered as a hard constraint. Owing to the fact that the severity target was thus prioritized, the system operational costs became a less critical factor in the decision making process and were consequently increased.

\begin{table}
\centering
\caption{Comparison of Cost Components (\euro)}
\label{breakdown_costs}
\begin{tabular}{cccc}
\sphline
&Preventive Cost&$\mathbb{E}$\{Corrective Cost\}&$\mathbb{E}$\{Severity\}\\
\sphline
Case A&2325&0.55&14.7\\
Case B &2650&0.03&9.3\\
\sphline
\end{tabular}
\end{table}

\subsubsection*{On the Utility of the Probabilistic Chance Constraint}

As cases A and B have exemplified,  security management involves trade-offs between  monetary and  societal costs. We recall that in case A, a smaller operational cost was achieved at the expense of neglecting the (low-probability) likelihood of very high severity levels. Arbitrating between high consequence, lower probability situations and the corresponding security costs has thus far been implicit under the N-1 criterion. The present case study serves to demonstrate that the proposal of this paper allows for an explicit arbitrage between these factors.

\begin{table}
\centering
\caption{Comparison of Alternative Chance Levels}
\label{chance_tab}
\begin{tabular}{cccc}
\sphline
\multicolumn{4}{c}{$P_g^0$  \hspace{2mm} (MW)}\\
\sphline
$\epsilon \setminus g$&1&2&3\\
\sphline
1&45&10&45\\
2&77.5&10&12.5\\
\sphline
\end{tabular}

\end{table}

To that end, we reconsider the example of figure \ref{fig:3bus} under two different upper bounds on the severity violation probability. A value of $\epsilon_1=10^{-5}$, \textit{i.e.} lower than the probability of realization of the high severity levels of case A, and a value of  $\epsilon_2=10^{-2}$. As table \ref{chance_tab} demonstrates, the solution to the former case   corresponds to the solution of case B. This is due to the fact that the probability of realization of the combination of contingencies and corrective control behaviors is greater than the upper bound on the severity violation probability. As such, the potential severity levels for these system states are to be controlled. Likewise, the solution to the case where $\epsilon_2=10^{-2}$ is the solution of case A since the severity threshold can be violated with a probability greater than the probability of realization of these states. In this manner, the present formulation allows for selecting the set of events against which the system should be protected by quantifying the relative costs and benefits.

\subsubsection*{On Relaxing the Line Capacity Limits}

As a final point for consideration in this set of case studies, we return to the fact that the proposal of this paper allows for a relaxation of the post-contingency line capacity limits. The sole criterion for allowing or prohibiting such a decision remains the resulting societal cost. This property has been demonstrated between cases A\footnote{We recall that this case is identical to the full formulation when the maximum admissible severity level has a considerably large value.} and B. In the former, line capacity limits under the corrective control failing behavior have been relaxed at the optimal solution. This was a result of the fact that the corresponding severity levels were admissible according to the respective threshold and violation probability upper bound. As in the latter this was not the case, line capacity limits in the corrective control failing behavior were not relaxed at the optimal solution.

\begin{table}
\caption{Case B: Preventive Dispatch with Potential Capacity Limit Relaxation ($MW$) }
\label{ALTcaseBprev}
\centering
\begin{tabular}{ccccc}
\sphline
$g$&1&2&3\\
\sphline
&45&10&45\\
\sphline
\end{tabular}
\vspace{+0.1in}
\caption{Case B: Corrective Re-dispatch with Potential Capacity Limit Relaxation ($MW$) }
\label{ALTBcaseBcorr}
\begin{tabular}{ccccc}
\sphline
$c\setminus g$&1&2&3\\
\sphline
2&55&10 & 35\\
3&45& 10& 45\\
4&45& 10& 45\\
5&x  & 50&  50\\
6&82.5& x &17.5\\
7&65& 35&   x\\
\sphline
\end{tabular}
\end{table}

In order to exemplify the efficiency of this approach with clarity, we conclude by presenting the solution to case B when the line capacity limits are allowed to relax even under the corrective control working behavior, tables \ref{ALTcaseBprev} and \ref{ALTBcaseBcorr}. As these tables demonstrate, this solution is identical to the initial solution of case B, since the relaxation these limits at the optimal solution would result in non-admissible severity levels. The fact that the identification of the conditions under which these limits can be relaxed is efficient, justifies the consideration of this additional degree of freedom to the decision maker.

\section{Multi-Area Case Study}

In the second set of case studies, we focus on the stakes of implementing the proposed approach across several areas of an interconnected power system. The scope of these case studies is to identify the combined effects of different security management practices on all end-users, located in different control areas of the system.

\begin{figure}
\centering
\includegraphics[width=0.43\textwidth]{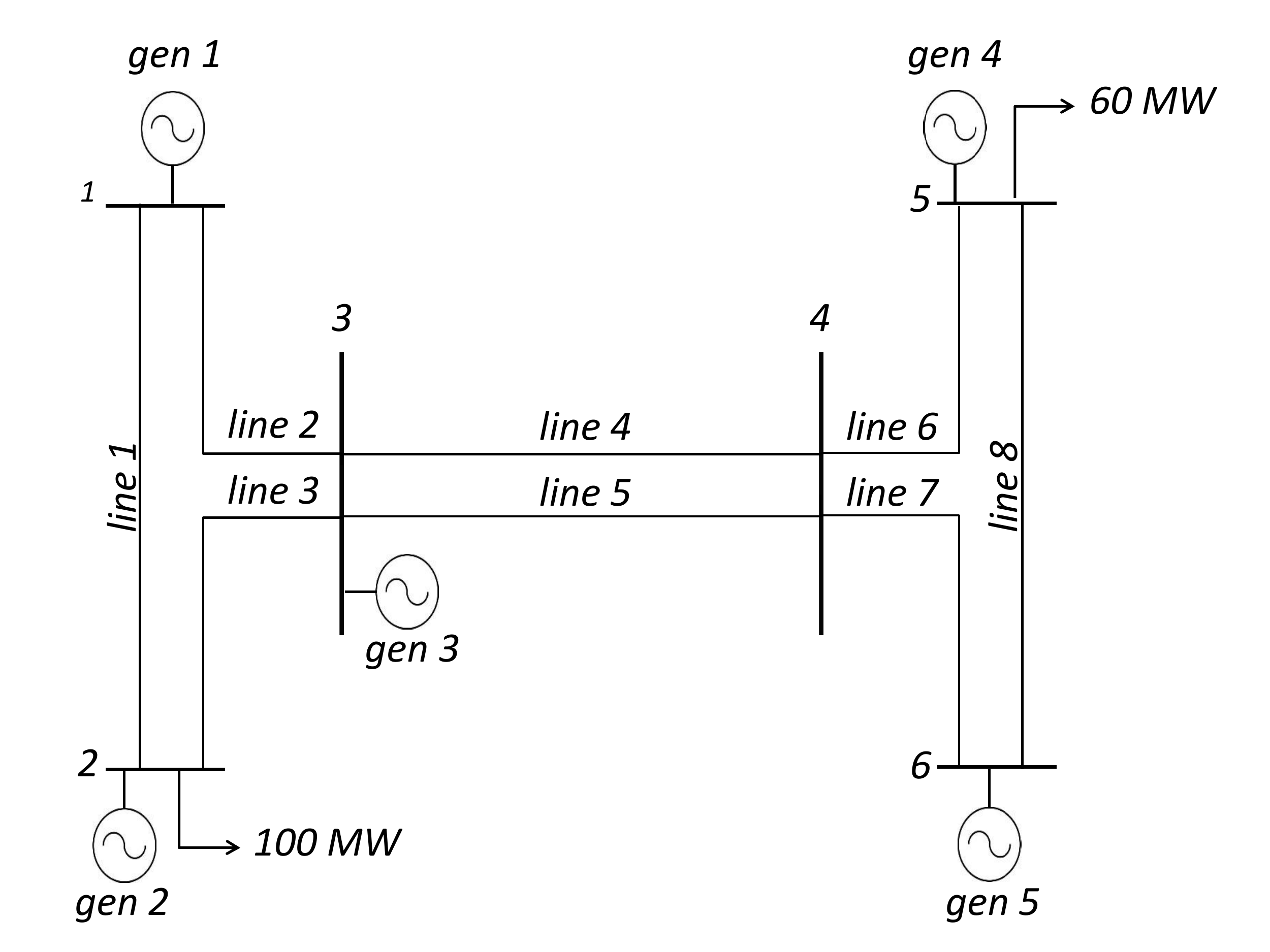}
\caption{Six-node, two area system}
\label{fig:twoareaz}
\end{figure}

\begin{table}
\caption{Generating Units Technical Characteristics }
\centering
\def~{\phantom{0}}
\begin{tabular}{ccccccc}
\sphline
\multirow{2}{*}{$g$} & $P_g^{min}$& $P_g^{max}$&$P_g^{-}$
&$P_g^{+}$ &$\Delta P_g^{e}$&MTTF\\
& \multicolumn{5}{c}{( MW)}&h)\\
\sphline
1 &  10&100&40&40&5&500\\
2 &  10&100&20&40&5&500\\
3 &  10&50&40&40&5&250\\
4 &  100&10&40&40&5&250\\
5 &  40&5&20&20&5&250\\
\sphline
\end{tabular}
\label{table:gens_twoarea_tech}
\vspace{+0.1in}
\caption{Generating Units Cost Data }
\begin{tabular}{cccc}
\sphline
\multirow{2}{*}{$g$}& $c_{g}$ & $c_{g}^r$ & $w_g$\\
& \multicolumn{2}{c}{(\euro/MWh)}&(\euro)\\
\sphline
1 & 20 & 5 &2500\\
2 & 40 & 8 &2500\\
3 & 30 & 7 &2500\\
4 & 65 & 10 &4000\\
5 & 50 & 8 &4000\\
\sphline
\end{tabular}
\label{table:gens_twoarea_costs}
\end{table}

To that end, we consider the six-node, two area system presented in figure \ref{fig:twoareaz}. Area A comprises  nodes (1-3) while nodes (4-6) form area B. The specifications of the 5 generating units are presented in tables \ref{table:gens_twoarea_tech} and \ref{table:gens_twoarea_costs}. Table \ref{twoarea_lines} lists the maximum capacity of the transmission lines, which are again assumed to have an equal admittance and a mean time to failure of 10000 hours. Finally, table \ref{sev_coeffs} shows the allocation of the severity coefficients across the two areas of the system. As this table indicates, area A is assumed to be of lower per unit severity (both in terms of shedding load and disconnecting generating units) with respect to area B.

\begin{table}
\centering
\caption{Transmission Capacity ($MW$)}
\label{twoarea_lines}
\begin{tabular}{cc}
\sphline
$\ell$&$f_{\ell}^{max}$ \\
\sphline
1&65\\
2 -- 7&55\\
8&65\\
\sphline
\end{tabular}
\end{table}

\begin{table}
\centering
\caption{Severity Coefficients}
\label{sev_coeffs}
\begin{tabular}{ccc}
\sphline
Area &$v_d$ (\euro/$MWh$)&$w_g$ (\euro)\\
\sphline
\multirow{2}{*}{A} &$150$& $2500$\\
&	$\{d=1\}$& $\{g\in[1,3]\}$ \\
\sphline
\multirow{2}{*}{B} &$300$& $4000$\\
&	$\{d=2\}$& $\{g\in[4,5]\}$ \\
\sphline
\end{tabular}
\end{table}

\subsection{System-wide Security Management}

In order to set a basis for comparison, we begin by considering that both areas are under the control of a common operator. In this context, we compare the effect of the proposed severity controlled approach against the N-1 across the two system areas. The set of credible contingencies includes the single failure of any system component.  As in the previous section, we assume that the probability of corrective control failure is equal to $0.2$. The admissible severity threshold is set at $25 \%$ of the maximum severity across both areas (\textit{i.e.} $s_{max}= 12125$ \euro). Finally, we consider this threshold as a hard constraint by setting the respective violation probability allowance to zero.

\begin{table}
\centering
\caption{System-wide Security Management Severity Levels(\euro)}
\label{table:system_wide}
\begin{tabular}{cccc}
\sphline
\multicolumn{4}{c}{N-1 Approach}\\
\sphline
Event& Total & Area A & Area B  \\
Line 1/Control Fail.&	19750	&13750 (69.6\%)&	  	6000 (30.4\%)	\\
Line 2/Control Fail.&	16750&16750  &	0\\
Line 3/Control Fail.&	19750	&13750 (69.6\%)&		6000 (30.4\%)	\\
Gen 1/Control Fail.&	14250	&14250 &		0\\
Gen 2/Control Fail.&	3750	&3750	 &	0   \\
Gen 3/Control Fail.&	1500	&1500	 &	0\\
Gen 4/Control Fail&	750	&750	&	0\\
\sphline
\multicolumn{4}{c}{ Severity Controlled Approach}\\
\sphline
Event& Total & Area A & Area B  \\
Gen 1/Control Fail.&	8250	&8250	&0\\	
Gen 2/Control Fail.&	6750	&6750	&0\\	
Gen 3/Control Fail.&	6750	&6750	&0\\	
Gen 4/Control Fail.&	1500	&1500	&0\\	
Gen 5/Control Fail.&	750	&750	&0\\	
\sphline
\end{tabular}
\end{table}

Table \ref{table:system_wide} analytically compares the potential severity levels between the N-1 and the proposed severity controlled approach\footnote{For the sake of the presentation simplicity, only events with non-zero severity values are henceforth listed.}. As anticipated, the potential severity levels  are greater under the N-1 approach. The allocation of the potential severity levels between the two areas is also of interest. Under the severity controlled approach, impacts to the end-users are always restricted within the low per unit severity area (area A). Nevertheless, this is not the case for the N-1 approach. We recall that under the N-1 approach the potential failure of corrective control is completely neglected. As such, there exist possible combinations of preventive control actions and corrective control behaviors under which  service disruption within the high per unit severity area is inevitable.

\subsection{Area-wide Security Management}

In the present subsection we consider the case wherein the two interconnected areas A and B are under the responsibility of different operators. Moreover, we consider that the operator of area A follows the proposed severity controlled approach while the operator of area B follows the N-1 practice. In the former occasion, the admissible severity threshold is set at 25\% of the maximum severity within area A and the violation probability allowance is once again equal to zero.

In order to evaluate the minimum system wide severity levels we adopt the following process:

\begin{itemize}
\item{We consider two separate security management subproblems concerning areas A and B. In both cases, the respective operator can only apply preventive and corrective actions  (and face the associated costs) within its area of interest. Moreover, the set of credible contingencies involves any single failure within the area under the responsibility of the respective operator as well as the failure of any single area interconnector.}

\item{ Given that the solution of subsection 2.1 only covers single failures, we assume that the occurrence of a contingency outside of the respective operator's control area is not a credible event. As such, we set the operating point of the generating units in the area outside the control of the respective operator according to the solution of an OPF concerning the two-area interconnected system, table \ref{table:OPF}. }

\item{Within the security management subproblem of operator A (adopting the severity controlled approach) we take into account the fact that it's jurisdiction does not extend outside area A. On this basis, we prevent the relaxation of any constraints within area B under any possible system state.}

\item{Within the security management subproblem of operator B (adopting the N-1 approach) we take into account the fact that the neglected potential failure of corrective control may cause constraint violations within area A as well.}

\item{Following the solution of subproblems concerning areas A and B we merge all identified corrective and preventive decisions and seek to minimize the net severity level across both areas under any credible state by means of emergency control actions.}

\end{itemize}

\begin{table}
\centering
\caption{System-wide Optimal Power Flow ($MW$)}
\label{table:OPF}
\vspace*{-2mm}\begin{tabular}{ccccc}
\sphline
            $P_1^0$&$P_2^0$&  $P_3^0$&$P_4^0$&  $P_5^0$\\
\sphline
           100&10&35&10&5\\
\sphline
\end{tabular}
\vspace*{-2mm}
\end{table}

\subsubsection{Severity Controlled Approach within Area A}

The scope of operator A is to optimize the value of \eqref{of} while maintaining the potential severity levels within area A below \euro 5625   (\textit{i.e.} $25 \%$ of the maximum possible severity within area A). By inspecting rows 8-10 of table \ref{table:system_wide}, which concern contingencies within area A, we can identify that the system wide security management strategy fails to achieve this target. We underline that all concerned instances are under the failure of corrective control. As such, operator A should seek for an alternative preventive dispatch to restrict the impact of these events. As demonstrated in table \ref{table:sys_vsA_disp}, the preventive actions selected by operator A  differ with respect to the case where the severity controlled approach is implemented across both system areas.

\begin{table}
\centering
\caption{Preventive  Dispatch Comparison ($MW$)}
\label{table:sys_vsA_disp}
\vspace*{-2mm}\begin{tabular}{cccccc}
\sphline
 Control          & $P_1^0$&$P_2^0$&  $P_3^0$&$P_4^0$&  $P_5^0$\\
\sphline
System&55&45&45&10&5\\
Area A&60&70&45&10&5\\
\sphline
\end{tabular}
\vspace*{4mm}
\caption{Severity Levels (\euro)}
\label{table:areaAsevs}
\vspace*{-2mm}
\begin{tabular}{cccc}
\sphline
\multicolumn{4}{c}{ Severity Controlled Approach in Area A}\\
\sphline
Event& Total  (\euro)& Area A & Area B  \\
Gen 1/Control Fail.& 18000 &0&18000	\\	
Gen 2/Control Fail.&	19500	&1500&18000	\\
Gen 3/Control Fail.&	4500&0&4500	\\
Line 1/Control Fail.&	20500&2500&18000	\\
\sphline
\end{tabular}
\vspace*{-2mm}
\end{table}

Evidently, the change in the preventive dispatch also gives rise to a change in the corrective re-dispatch.  Table \ref{table:areaAsevs} demonstrates the optimal severity levels from the perspective of area A operator.  We note that rows 1-3 of  \ref{table:areaAsevs}  correspond to rows 8-10 of table  \ref{table:system_wide}. All these events concern the failure of corrective control following the occurrence of a generating unit outage. The corresponding non-zero severity levels of these events denote the necessity of emergency load shedding across the system. In the present case, transmission network constraints are not restrictive to impose whether load should be shed within area A or area B to alleviate the mismatch. Under these conditions, from the perspective of operator A, the acceptable solution would be to transfer the impact of the energy mismatch to area B, by prioritizing the disruption of the demand connected at node 5.

%\begin{table}
%\centering
%\end{table}

\begin{figure}
\centering
\includegraphics[width=0.4\textwidth]{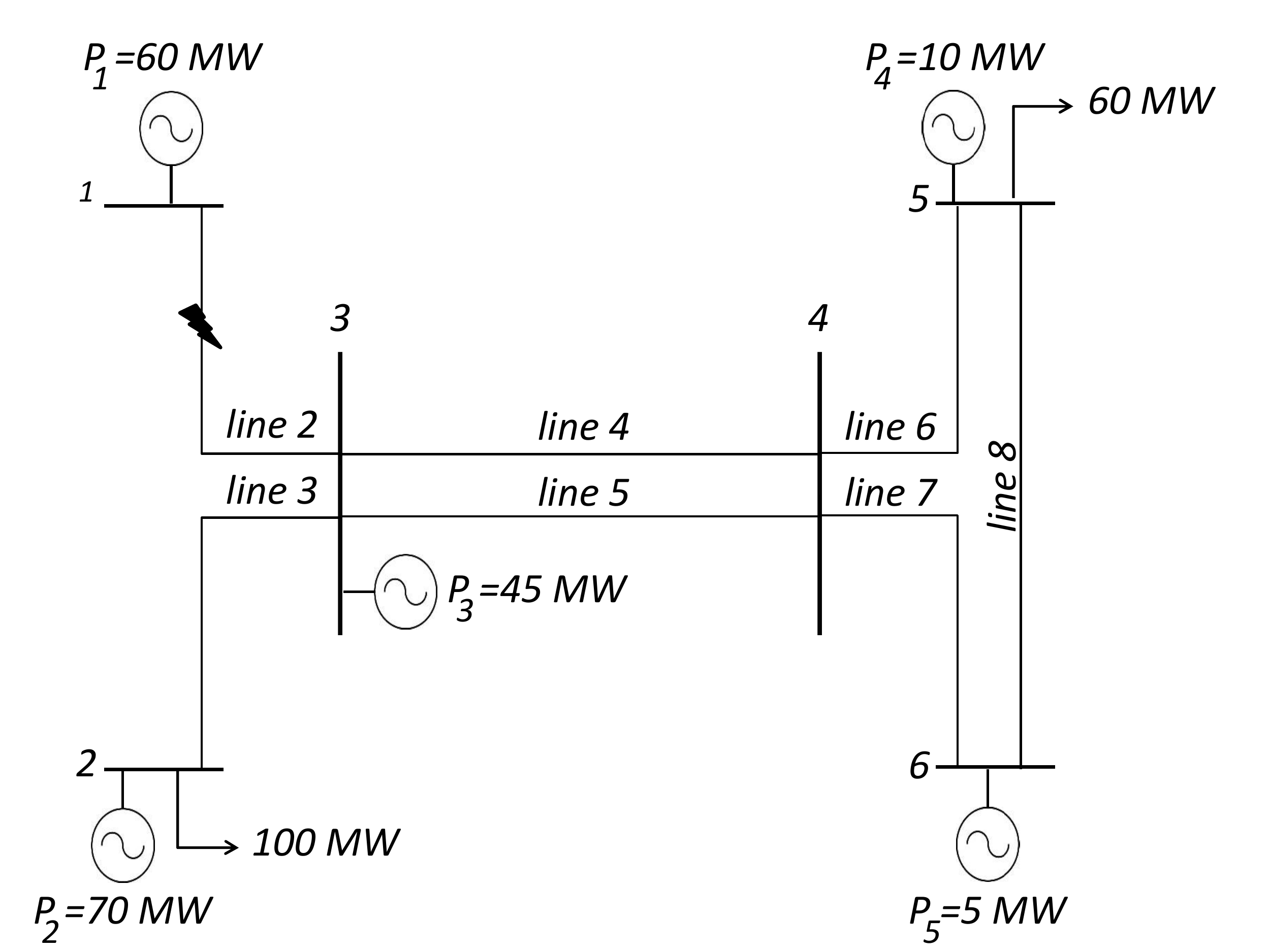}
\caption{Failure of Corrective Control under Outage of Line 1}
\label{fig:twoareaz_failure}
\end{figure}

In order to exemplify the values listed in the last row of \ref{table:areaAsevs}, we present in figure \ref{fig:twoareaz_failure} the network configuration following the outage of line 1 and the failure of corrective control. We recall, that in such an occasion all generating units would remain at the operating points listed in the second row of table \ref{table:sys_vsA_disp}. Consequently, the 60 MW of injection at node 1 would result in the overload of line 2, which has a maximum capacity limit of 55 MW. The loss of this line would result in an  deficit of 60 MW. Considering, once again that such a deficit can be alleviated by shedding the load at node 5 operator A deems this event acceptable. In the event that the load at node 5 is shed, the sole impact in area A would be the disconnection of unit 1 at a severity of \euro 2500.

\subsubsection{ N-1 Approach within Area B}

For the sake of completeness, we briefly discuss the adoption of the N-1 approach within area B. By inspecting table \ref{table:gens_twoarea_tech}   we can identify that the minimum stable restrictions of generating units 4 and 5 prohibit any reconsideration of the operating points listed in table \ref{table:OPF}. Considering the network topology (figure \ref{fig:twoareaz}) as well as the line capacity limitations (table \ref{twoarea_lines}), such operating point is feasible under any single line outage within area B, as well as the outage of  any single interconnector. As a result, the N-1 approach in area B would only involve corrective actions in the case  any one of the two generating units trips.

\subsubsection{The Combined Effect of the Different Security Management Practices}

We conclude  by assessing the combined effect of implementing different security management approaches in the two areas of the interconnected system of figure \ref{fig:twoareaz}. For this purpose, we re-compute the system-wide optimal severity levels in the case where the preventive and corrective strategies for areas A and B have been identified according to the severity controlled approach and the N-1 approach respectively.

\begin{table}
\centering
\caption{ Minimum System-wide Severity Levels (\euro)}
\label{table:system_min_sev}
\begin{tabular}{cccc}
\sphline
Event& Total&Area A & Area B \\
\sphline
Line 1/Control Fail.&	11500&11500&0	\\
Gen 1/Control Fail.& 9000 &9000&0	\\	
Gen 2/Control Fail.&	10500	&10500&0	\\
Gen 3/Control Fail.&	2250&2250&0	\\
Gen 4/Control Fail.&	1500	&1500	&0\\	
Gen 5/Control Fail.&	750	&750	&0\\	
\sphline
\end{tabular}
\end{table}

As table \ref{table:system_min_sev} demonstrates, in order to minimize the system-wide severity levels the impact of any event should be once again restricted within area A. We highlight the fact that the potential severity levels in area A are considerably greater with respect to the values listed in table \ref{table:system_wide}. Evidently, this is due to the fact that the actual criticalities of the end-users located at area B where
completely neglected by area A operator under the severity controlled approach.

\begin{table}
\centering
\caption{Area A Security Management Costs (\euro)}
\label{operA_costs}
\begin{tabular}{cccc}
\sphline
Control &Preventive Cost&$\mathbb{E}$\{Corrective Cost\}&$\mathbb{E}$\{Severity\}\\
\sphline
System&4250.5&-1.58&12.44\\
Area A&4450&-0.318&10.88\\
\sphline
\end{tabular}
\end{table}

Table \ref{operA_costs} presents a comparison of the cost breakdown in area A between the case wherein the severity controlled approach was adopted in both areas of the interconnected system and the case where the severity controlled approach was adopted only in area A.  Recalling that the need to control the severity under corrective control failure drives the operator towards more expensive preventive control strategies justifies the observed increased in the preventive cost. In the system-wide case a larger severity threshold was available to the operator. We observe once again from table \ref{table:system_wide} that such large threshold allowed the severity in area A to reach values above $25\%$ of the maximum possible area A severity. As in the case where severity control is only implemented in area A the $25\%$ of the maximum area severity must be enforced, a more conservative, hence more expensive, preventive dispatch has been selected.

In both cases, the negative terms related to the expected corrective control costs highlight the push toward the more conservative preventive control strategies. These negative terms denote that the preventive schedules are so conservative that if a failure indeed triggers the operation of corrective control system costs can be reduced. We should finally clarify the origin in the observed reduction in the expected severity value. Even though the severity levels listed in table \ref{table:system_min_sev} exceed those of table \ref{table:system_wide}, the considerable reduction in case that corrective control has failed following the outage of unit 3, along with the relatively large probability of realization of this event, lead to a reduced expected value.

Table \ref{operB_costs} introduces a similar comparison for Area B. As already noted, the preventive dispatch remains unaltered in the case that area B follows the N-1 approach and the case where the severity controlled approach is implemented in the two system areas. Moreover, in both cases the potential severity levels would be equal to zero since the impact of any event would be directed toward the lower (per unit) severity area A. The increase in the expected corrective costs in the system-wide severity controlled case arises from the fact that generating units in area B are re-dispatched to counteract contingencies in area A.

\begin{table}
\centering
\caption{Area B Security Management Costs (\euro)}
\label{operB_costs}
\begin{tabular}{cccc}
\sphline
Control &Preventive Cost&$\mathbb{E}$\{Corrective Cost\}&$\mathbb{E}$\{Severity\}\\
\sphline
System&900&1.165&0\\
Area B&900&0.47&0\\
\sphline
\end{tabular}
\end{table}

\section{Concluding Remarks}

The N-1 practice has been pivotal to the secure operation of electrical power systems. On the core of this approach, lies a rigid attitude with regard to the uncertainties in power system operation. Events that are considered sufficiently likely (\textit{i.e.} the outage of any single system component) are to be treated as threats to the system security whereas every other event is to be neglected. The vast experience from the operation of power systems to date has exemplified the resilience of this approach  in the case the degree of uncertainty is relatively minor. Nevertheless, even under such case, the rare examples where this approach failed to protect the system have resulted in severe consequences to its end-users. This is due to the fact that the N-1 approach inherently disregards the difference in the risk levels between the different possible system operating points \cite{Daniel_risk}.

Nowadays, the growth of uncertainties is a well-acknowledged fact amongst the members of the power system  community.  To cater for this fact under the N-1 framework in an efficient manner, more and more corrective control measures have to be adopted. The increasing adoption of such measures may not fully warrant the security of modern power systems though. Understanding the behavior of corrective control and, most importantly, characterizing the uncertainties induced by its operation appear as preconditions to this task. In a different case, the use of corrective control measures would introduce hidden threats of potentially high consequences to the system security.

Considering these facts, in the present paper we investigated the utility of an alternative probabilistic security management approach. This approach departs from the N-1 practice in the following key aspects: i) the probability of realization of any credible system state is explicitly taken into account, ii) corrective control outcome is considered as an additional source of uncertainty in the system operation, iii) the  potential consequences  resulting from any combination of contingencies and corrective control outcome, both in terms of operational costs to the TSO and in terms of service end-users' interruption costs, serve as a criterion for the treatment of possible events as threats to be counteracted and iv) the possibility of inducing severe consequences to the end-users is explicitly controlled, by means of a probabilistic chance constraint.

We would like to stress that the proposed probabilistic framework is a {\em proper} generalisation of various deterministic (preventive/corrective) SCOPF formulations used for security management in practice,  in the sense that one can obtain these latter by setting $s_{\max}$ and $\epsilon$ to zero, by leaving out the expected cost of service interruption from the objective function, and by assuming particular choices of the corrective control behaviors.   We also remark that, while this framework is designed as a tool for {\em designing} optimal combined preventive and corrective control strategies for real-time operation, it may as well be used as a tool for {\em evaluating} any other  (partially or completely specified) alternative control strategy by translating it in the form of additional (hard, or possibly soft) constraints and incorporating these latter into the RTP formulation. These two characteristics enable, in particular, a fair comparison of the current deterministic security management criteria with the proposed probabilistic one, which is a necessary condition for gaining acceptance and enabling the migration  in real-world practice.

Section \ref{section:math} furthermore provides a detailed mathematical formulation of the proposed framework, under the DC approximation and some additional simplifying assumptions, yielding a MILP problem which can be efficiently solved with available tools, even for large-scale systems.

\subsection{Main Findings}

Albeit remaining at the \textit{proof-of-concept} level, the initial findings of this work (demonstrated by example in sections IV and V) clearly establish the interest of pursuing this research direction.

The single-area case studies analyzed in section IV have, first and foremost, unveiled the limitations in the scope of the N-1 approach. In contrast, these case studies have demonstrated the greater potential of a probabilistic, severity controlled approach. By means of this analysis, we have established that the proposed approach is in principle efficient in arbitrating between the operational costs of security provision and the potential consequences of service interruption to the system end-users. Moreover, we have shown the fitness of achieving a probabilistic guarantee with respect to mitigating high consequence threats of low likelihood. Not to be neglected, the outcome of these case studies accords to the inevitable truth: achieving an enhanced security level comes at (possibly, dis-proportionally) greater monetary costs. It thus appears of critical importance to consider both factors in order to develop an efficient security management strategy.

The multi-area case studies of section V considered the effect of adopting two different security management practices within two sub-areas of an interconnected system. The findings of these case studies serve to raise the need for conformity in the practices adopted by TSOs within an interconnected system.  We have demonstrated that without such conformity, a single TSO may fail to achieve the severity related targets within his own area of interest even in the case wherein his individual policy appears effective in doing so. Given that the technical interconnection remains a physical reality, a `myopic' perspective within a sub-area of the system may well be counter-productive. It follows that the efficient implementation of such a framework calls for additional information exchange amongst the TSOs in an interconnected system. As this case study exemplified, in order to make sound decisions, any single TSO must be accurately informed on the factors driving emergency control (\textit{i.e.} range of controllability and criticalities of service disruption) in the system areas outside its jurisdiction. 

\subsection{Open Issues}

The present work serves to establish the interest in migrating from the N-1 practice toward a probabilistic security management approach. We thus conclude by discussing a series of issues that remain open in the direction of realizing such a transition.

\subsubsection{Characterizing the Behavior of Corrective Control}

The unpredictability in the behavior of corrective control was adopted as a starting point for this work. Given the role of corrective control, even in today's power system security management, characterizing the behavior of this resource with high accuracy would be of critical importance. In the context of the proposal of this paper, further research would be required to identify an exhaustive set including all the possible failing behaviors of corrective control. Evidently, associating a credible probability value to any such failing behavior is also essential.

In addition to developing such a set, special care is required in order to accurately model the operation of emergency control. The scope for this activity would be to reflect all the dynamic phenomena  occurring in subsequence to the realization of any possible corrective control failing behavior. We underline that achieving a high degree of accuracy while preserving the computational tractability of the decision making problem is a non-trivial task. With the intention of concentrating on the fundamental principles of the proposed security management practice in this work,  less emphasis was placed thus far on the former. Given the criticality of this point on the proposed decision making framework, in the next stage of this research we will seek to identify an optimal trade-off between computational tractability and modeling accuracy.

\subsubsection{Quantifying the Societal Cost of Service Interruptions}

Evaluating the utility received by the electricity consumers has been an active field of research within the social sciences community (see, \textit{e.g.} \cite{Spain,Ireland}). Needless to say, this field lies well outside the scope of any engineering study. Be that as it may, we take this opportunity to highlight the gravity of this point in completing the migration from the N-1 approach to novel security management strategies.

 In the development stages of this work, we identified an additional societal cost component  related to the loss of generating units from the system. As presented in the earlier parts of this paper, for the purposes of security management in real-time, this cost component should encapsulate the effect of the loss of any generating unit on the continuity of supply further than the period of interest. Having established in this paper the scope for considering such a factor in the decision making process, we thus raise the necessity of accurately quantifying this value.

As already mentioned, the quantitative results presented in this paper are not based on validated reference values for the aforementioned cost components. To the best of the authors' comprehension,  this remains in complete accordance with the purpose of establishing the fundamental properties of the proposed framework.

\subsubsection{Developing Tractable Solution Algorithms for Large-Scale Power Systems}
In parallel to the issues discussed in the preceding paragraphs, we should re-state that the exact solution of the optimization problem considered in this paper in large-scale power systems is a challenging task in terms of computational complexity. In the next steps of this work we intend to consider:

\begin{itemize}
\item{Re-casting the problem under consideration in the AC power flow context. In this context, we aim not only to enhance the representation of the system behavior but also to extend the set of potential corrective control actions beyond the set adopted in the development of the present work. Recent works from the applied mathematics community, providing admissible convex relations of the AC OPF may show to be useful in this context \cite{Lavaei2012}.}

\item {Defining approximations of the cost function \eqref{obj} on the basis of the scenario tree approach from the multi-stage stochastic programming literature  \cite{defourny2012scenario}.}

\item Replacing the chance constraint \eqref{cc} by an appropriate number of hard constraints defined over a sample of scenarios  drawn according to the probabilistic model of our problem, building on the recent results given in  \cite{campi-garatti-2008,campi-garatti-2011}.

\item{ Identifying a well-chosen subset of the combinations of contingencies and  behaviors that have to be taken into account as credible events in order to warrant the system security level. We note that the starting point for the formulation of such a subset extends beyond the set of events considered under the N-1 approach (\textit{i.e.} the failure of any single system component) to encapsulate all the possible events according to the vulnerability of the system components. }

\item{Evaluating the performance of the proposed framework on a large-scale power system with respect to the N-1 approach, as implemented in today's practice by TSOs. To that end, we will exclude the consideration of the probability of occurrence of the various credible contingencies from the model of the N-1 approach. }

\end{itemize}

\subsubsection{Extension to Alternative Decision Making Horizons}

As a final point, we return to the multi-stage nature of the power system security management problem. The consideration of the latest decision stage in this work, serves as a first step toward developing a coherent approach across the full set of the overlapping decision making horizons. 

To that end, we suggest to gradually extend this research by working backwards, from real-time to intraday and day-ahead operation planning, then maintenance management, then system expansion. Notice that by doing so the space of uncertainties to be modelled and the space of decisions to be jointly optimised will have to be gradually expanded. 

In particular, as soon as we reach the operation planning stage, one major additional source of uncertainty to be accounted for are the power injections assumed from renewable generation, demand, and market driven dispatchable generation units. 

\section*{Acknowledgements}

This work is supported by the Belgian
Network DYSCO, an Interuniversity Attraction Poles Programme
initiated by the Belgian State, Science Policy Office. 
The authors are also most grateful to  Florin Capitanescu for his useful discussions and suggestions.

\section*{Appendix A}

The present appendix outlines the linearization of products of binary and continuous variables appearing in the formulation presented in section \ref{section:math}. More specifically, equalities \eqref{low_pf} and \eqref{low_pf2} involve products of binary variable $\lambda_{\ell}^c(b)$ and free continuous variable $\hat{\theta}_{n}^c(b)$ as in,

\begin{flalign}
&\hat{f}_{\ell}^c(b)- \frac{a_{\ell}^c \cdot (1-\lambda_{\ell}^c(b)) }{X_{\ell}}\sum_{n \in \mathcal{N}_n} \beta_{n,\ell} \cdot \hat{\theta}_{n}^c(b)   =0&\\
&\hat{f}_{\ell}^c(b)- \frac{a_{\ell}^c \cdot (1-\tau_c\cdot \lambda_{\ell}^c(b)) }{X_{\ell}}\sum_{n \in \mathcal{N}_n} \beta_{n,\ell} \cdot \hat{\theta}_{n}^c(b)   =0.&
\end{flalign}

The product $\lambda_{\ell}^c(b) \cdot \hat{\theta}_{n}^c(b)$ can be replaced by free continuous variable  $\tilde{\theta}_{n}^c(b)$  via the following set of constraints:
\begin{flalign}
&\tilde{\theta}_{n}^c(b)\le \lambda_{\ell}^c(b) \cdot  M&\\
&\tilde{\theta}_{n}^c(b)\ge -\lambda_{\ell}^c(b) \cdot  M&\\
&\tilde{\theta}_{n}^c(b)-\hat{\theta}_{n}^c(b) \le \left(1-\lambda_{\ell}^c(b)\right) \cdot M&\\
&\tilde{\theta}_{n}^c(b)-\hat{\theta}_{n}^c(b)\ge \left(\lambda_{\ell}^c(b) -1 \right)\cdot M. &
\end{flalign}

Inequalities \eqref{low_uppcoup} through \eqref{low_dwncoup2} involve products of binary variable $y_g^c(b)$ and non-negative continuous variables $P_g^0, \hspace{1mm} P_g^c$ as in,
\begin{flalign}
&\hat{P}_{g}^c(b) -  \left(1-y_g^c(b)\right) \cdot  P_g^c  \le 0&\label{low_uppcoup_app}\\
&-\hat{P}_{g}^c(b)+\left(1-y_g^c(b)\right)  \cdot\left( P_g^c  -a_{g}^c \cdot  \Delta P_{g}^{e}\right)  \le 0 &\label{low_dwncoup_app}\\
&\hat{P}_{g}^c(b) -  a_g^c \cdot \left(1-y_g^c(b)\right) \cdot  P_g^0  \le 0&\label{low_uppcoup2_app}\\
&-\hat{P}_{g}^c(b)+a_g^c \cdot \left(1-y_g^c(b)\right) \cdot \left( P_g^0  -  \Delta P_{g}^{e}\right)  \le 0.& \label{low_dwncoup2_app}
\end{flalign}
The product $y_g^c(b) \cdot P_g^0$ can be replaced by non-negative continuous variable $\tilde{P}_g^0 \ge0$ through\footnote{The same process is applied to replace the product $y_g^c(b) \cdot P_g^c$ by non-negative continuous variable $\tilde{P}_g^c \ge0$.}:
\begin{flalign}
&\tilde{P}_g^0 \le y_g^c(b)\cdot  M&\\
&\tilde{P}_g^0 -P_g^0\le \left(1-y_g^c(b)\right) \cdot M&\\
&\tilde{P}_g^0-P_g^0 \ge \left(y_g^c(b)-1 \right)\cdot M. &
\end{flalign}

\section*{Appendix B}

The MILP formulation of \eqref{of}-\eqref{chance_two} may be extended in a straightforward way to  piece-wise linear costs functions instead of the objective  \eqref{of}. In the present appendix we explain how this may be exploited to adopt a different objective function, according to an incremental settlement of preventive and/or corrective control costs.

\subsection*{Preventive Control Adjustment Cost}

In some regulatory jurisdictions, a TSO would only be liable for the cost of deviations with respect to the outcome of an already settled market clearing, denoted henceforth as $P_g^M$ for every unit $g \in \mathcal{N}_g$. In order to model this, in the general case wherein different fees are applicable to upwards or downwards changes, the first term of \eqref{of} would be modified in  the following way:
\begin{flalign}
&C_0(x_0,u_0)=\left(\sum_g \overline{c}_g \cdot \overline{P}_g^M+ \sum_g \underline{c}_g \cdot \underline{P}_g^M\right), & \label{appb.two}\\
\intertext{where symbols $\overline{c}_g$ and $\underline{c}_g$ denote the per unit costs of upwards and downwards deviations respectively and:} 
&\overline{P}_g^M=\max\left\{\left(P_g^0-P_g^M\right),0\right\}\label{appb.three}\\
&\underline{P}_g^M=\max\left\{\left(P_g^M-P_g^0\right),0\right\}\label{appb.four}.
\end{flalign}

%\intertext {where symbol $P_g^M$ denotes the outcome of the market clearing process for generating unit $g$. }
The expression of \eqref{appb.three} and \eqref{appb.four} in a linear form can be made by the use of a set of  logical constraints. We also notice that the case wherein deviations are remunerated according to an absolute value (\textit{i.e.} $|P_g^M-P_g^0|$) is a  particular case of \eqref{appb.two}-\eqref{appb.four} where $\overline{c}=\underline{c}_g$.

\subsection*{Corrective Control Adjustment Cost}

A commonly used alternative for the costs of corrective control penalises  any  modification against the preventive schedule. In a similar manner to \eqref{appb.two}-\eqref{appb.four}, this can be  integrated in the MILP of \eqref{of}-\eqref{chance_two} via:
\begin{flalign}
&C_c(x_0,u_0,c,u_c)=\left(\sum_g \overline{c}^r_g \cdot \overline{P}_g^c+ \sum_g \underline{c}^r_g \cdot \underline{P}_g^c\right),&\label{appb.five}\\
\intertext{where symbols $\overline{c}^r_g$ and $\underline{c}^r_g$ denote the per unit costs of incremental/decremental deviations respectively, while:} 
&\overline{P}_g^c=\max\left\{\left(P_g^c-P_g^0\right),0\right\}\label{appb.six}\\
&\underline{P}_g^c=\max\left\{\left(P_g^0-P_g^c\right),0\right\}.\label{appb.seven}
\end{flalign}

%\section*{Acknowledgements}

%{\small ... }

\bibliographystyle{ieeetr}
\bibliography{irep-2013}

\end{document}